\documentclass[11pt]{amsart}
\usepackage[english]{babel}
\usepackage{xcolor}
\usepackage{accents}
\usepackage{geometry}
\usepackage{graphicx, endnotes, setspace,  psfrag}
\usepackage{amssymb, dsfont, amsthm, amsmath, bm, mathrsfs, yfonts,natbib}
\newtheorem{thm}{Theorem}[section]

\newtheorem*{prop*}{Proposition}

\newtheorem{defn*}{Definition}
\theoremstyle{remark}
\newtheorem{rem}[thm]{Remark}
\newtheorem*{rem*}{Remark}
\newtheorem{example}[thm]{Example}
\newcommand{\set}[1]{\left\{#1\right\}}
\newcommand{\ccross}[3]{\begin{minipage}{50pt}\psfrag{a}[r]{$#1$}\psfrag{b}[c]{$#2$}\psfrag{c}[l]{$#3$}\includegraphics[width=50pt]{crosm}\end{minipage}}
\newcommand{\trr}{\triangleright}
\newcommand{\brr}{\blacktriangleright}
\newcommand{\rrt}{\triangleleft\,} 

\newcommand{\ass}{\stackrel{\textup{\tiny def}}{=}}

\newcommand{\bra}{\left|}
\newcommand{\ket}{\right \rangle}
 \newcommand{\dU}{ \, \raisebox{1.25pt}{\scalebox{.8}{$\coprod$}} \; }
\newcommand{\kett}{\left \langle}
\newcommand{\brat}{\right|}

\newcounter{saveenumerate} 
\makeatletter
\newcommand{\enumeratext}[1]{%
\setcounter{saveenumerate}{\value{enum\romannumeral\the\@enumdepth}}
\end{enumerate}
#1
\begin{enumerate}
\setcounter{enum\romannumeral\the\@enumdepth}{\value{saveenumerate}}%
}
\makeatother

\begin{document}

\title{Tangle Machines}

\author{Avishy Y. Carmi and Daniel Moskovich}

\address{Faculty of Engineering Sciences \& \\ Center for Quantum Information and Technology \\ Ben-Gurion University of the Negev, Beer-Sheva 8410501, Israel}

\date{3rd of March, 2015}



\begin{abstract}
Tangle machines are topologically inspired diagrammatic models. Their novel feature is their natural notion of equivalence. Equivalent tangle machines may differ locally, but globally they share the same information content. The goal of tangle machine equivalence is to provide a context-independent method to select, from among many ways to perform a task, the `best' way to perform the task. The concept of equivalent tangle machines is illustrated through examples in which tangle machines represent networks for distributed information processing, networks of adiabatic quantum computations, and iterative computations.
\end{abstract}

\label{firstpage}
\maketitle


\section{Introduction}

\subsection{The idea in a nutshell}

This paper introduces a diagrammatic formalism for computation and for information processing. Behind this endeavor is the observation that the combinatorial properties of knot diagrams mimic principles pertaining to conservation and to manipulation of information in networks. 
Our approach is low-dimensional topological, whereas previous diagrammatic descriptions of information flow in networks have been in terms of labeled graphs.

\begin{figure}
\centering
\includegraphics[width=0.9\textwidth]{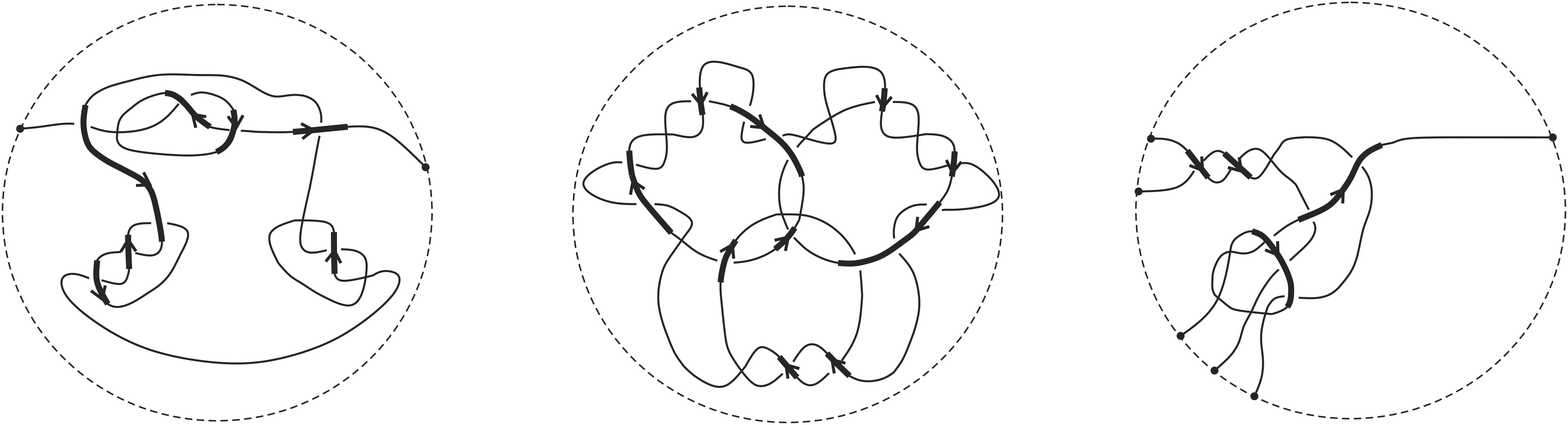}
\caption{\label{F:Machine} A tangle machine with colours suppressed.}
\end{figure}

We construct diagrammatic models called \emph{tangle machines}, or just \emph{machines} for short, represented by labeled versions of diagrams such as those of Figure~\ref{F:Machine}, that represent entities and relationships between those entities. Unlike labeled graphs, in which edge $e$ from vertex $a$ to vertex $b$ represents a transition from the label of $a$ to the label of $b$, the basic building block of a tangle machine is an \emph{interaction}, in which \emph{agent} $c$ causes a transition from colours of \emph{input patients} $a_1,a_2,\ldots,a_k$ to colours of corresponding \emph{output patients} $b_1,b_2,\ldots,b_k$. A tangle machine (which we call a \emph{machine} for short) makes explicit the \emph{cause} of a transition. From one perspective, a machine is a computational scheme, a sort of ``planar algorithm'' wherein interactions represent basic computations. From the dual perspective, a machine is a network within which information is manipulated at interactions and then transmitted further down to registers at other interactions. Information can be both a patient (\textit{e.g.} an input data stream) and an agent (\textit{e.g.} commands of a computer programme). This aspect of information is captured by tangle machines but not necessarily by labeled graphs.

The novel feature of tangle machines is their flexibility. Whereas competing graphical models are rigid, tangle machines admit a natural local notion of \emph{equivalence}. Roughly speaking, two machines are equivalent if one can be perfectly reproduced from the other. Machine equivalence parallels the notion of \emph{ambient isotopy} in low dimensional topology. We consider a `crossing' (which we call an \emph{interaction}) to represent a computation (in the sense of computer science or of automata) or a fusion of information whose basic symmetries are encapsulated by the three \emph{Reidemeister moves} of Figure~\ref{F:local_moves_machines1} in Section~\ref{S:Machines}\ref{SS:Equivalence}. Topology suggests that these three local rewrite moves are in a sense different aspects of a single operation consisting of rotating a plane onto which an embedded object is projected.

Local features such as implementation and performance of computations or of information manipulations modeled by the tangle machine may be different for networks modeled by equivalent machines, but we consider the global \emph{information content} of such networks to be the same. We may thus use the tangle machine formalism to select, from among many equivalent models which `perform the same task', the model (and thus the network) best suited for a specified application. This concept is illustrated in our examples:

\begin{itemize}
\item Machines representing networks of distributed information processing (Section~\ref{S:information}).
\item Machines representing adiabatic quantum computations (Section~\ref{S:AQC}).
\item Machines representing iteration and Markov chains (Section~\ref{S:Recursion}).
\end{itemize}

 All of the examples make use of the tautological fact that colours of endpoints are \emph{machine invariants}. In each example, in order to illustrate the operational meaning of machine equivalence, three equivalent machines with different local features are examined.

Our paper is organized as follows. In the introduction we give a toy example roughly explaining what a tangle machine is, followed by a summary of our examples. In Section~\ref{SS:Context} we discuss several aspects of how tangle machines fit into a wider scientific context. In Section~\ref{S:Machines} we define tangle machines and tangle machine equivalence, deferring technical details to the appendix. Following this are our examples, which can be read independently from one another, followed by the conclusion.

\subsection{What is a (tangle) machine?}

\begin{figure}[htb]
\centering
\psfrag{a}[c]{\small $\hat{Y}_t$}
\psfrag{b}[c]{\small $\hat{X}_t$}
\psfrag{c}[c]{\small $\hat{Z}_t$}
\psfrag{e}[c]{\small $\brr$}
\psfrag{e1}[c]{\small $\trr$}
\psfrag{x}[l]{\small $\hat{X}_t \brr \hat{Z}_t$}
\psfrag{y}[r]{\small $\hat{X}_t \trr \hat{Y}_t$}
\psfrag{d}[l]{\small $\hat{Y}_t \brr \hat{Z}_t$}
\psfrag{f}[r]{\small $\left(\hat{X}_t \trr \hat{Y}_t\right) \brr \hat{Z}_t$}
\psfrag{g}[r]{\small $\left(\hat{X}_t \brr \hat{Z}_t\right) \trr \left(\hat{Y}_t \brr \hat{Z}_t\right)$}
\includegraphics[width=0.7\textwidth]{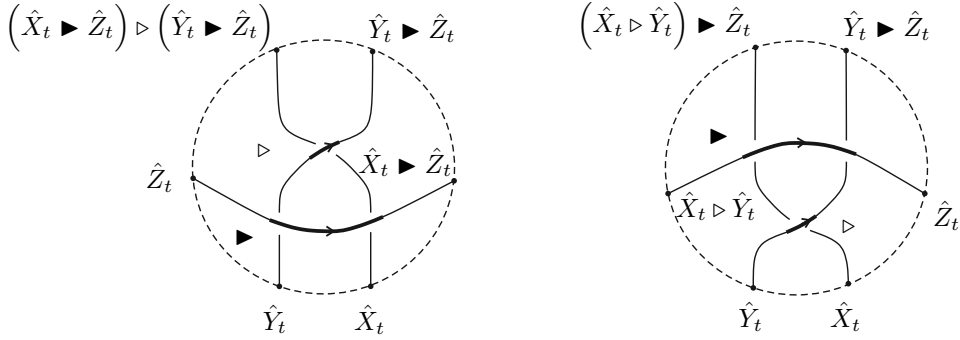}
\caption{\label{F:toy} Two equivalent data stream fusion networks described by two equivalent tangle machines.}
\end{figure}

To give some idea of the sorts of things that tangle machines can be useful for, consider the following toy example.
Three transmitting devices $x$, $y$, and $z$ continuously stream data $\hat{X}_t$, $\hat{Y}_t$, and $\hat{Z}_t$. Our task is to combine these data streams, eliminating redundancy (\textit{e.g.} because of the Problem of Double Counting \citep{Jazwinski:70}). Two schemes to combine the data streams are represented by the machines in Figure~\ref{F:toy}.

In the left machine, combine $\hat{X}_t$ and $\hat{Y}_t$ with $\hat{Z}_t$ to obtain fused data streams $\hat{X}_t\brr \hat{Z}_t$ and $\hat{Y}_t\brr \hat{Z}_t$. We indicate that these two fusions are performed independently by thickening the strand labeled $\hat{Z}_t$.
We read the fusion from bottom to top because the overcrossing arc is directed from left to right. If it were oriented from right to left, we would read from top to bottom and we would be filtering out the data stream $\hat{Z}_t$ from $\hat{X}_t\brr \hat{Z}_t$ and from $\hat{Y}_t\brr \hat{Z}_t$. At the top, $\hat{X}_t\brr \hat{Z}_t$ is combined with $\hat{Y}_t\brr \hat{Z}_t$ using a possibly different operation $\trr$ to combine the data streams.

 The left and the right machine describe equivalent data stream fusion schemes, which is visually indicated by the fact that they are related by sliding one overcrossing arc over another. Indeed, the result of the data stream fusion in the right machine is $\left(\hat{X}_t\trr \hat{Y}_t\right)\brr \hat{Z}_t$, which is the same combined data stream as in the left machine because redundant double~appearance of $\hat{Z}_t$ in $\hat{X}_t\brr \hat{Z}_t$ and $\hat{Y}_t\brr \hat{Z}_t$ is eliminated by $\trr$. This redundancy elimination is possible only because we know that $x$ and $y$ independently fused their data streams with $z$.

 However, there is an important difference between these two schemes in Figure~\ref{F:toy}. Imagine that, at some time $t_1>0$, device $y$ becomes faulty. In this case, the left machine is superior because it contains the intermediate data stream $\hat{X}_t\brr \hat{Z}_t$ which might be useful even when $\left(\hat{X}_t\brr \hat{Z}_t\right)\trr \left(\hat{Y}_t\brr \hat{Z}_t\right)$ is junk. Conversely, if $z$ becomes faulty at some time $t_2>0$, then the right machine would be preferred. The top overcrossing arc might slide back and forth at different times. Thus, the machines in our toy example might be describing the underlying logic of a simple self-optimizing fault-tolerant data stream fusion network.

 These ideas will recur in various contexts throughout the paper:

 \begin{enumerate}
 \item Reversibility and redundancy elimination hardwired into the formalism (see Section~\ref{S:Machines}\ref{SS:Quandle}).
 \item Hardwired independence (the thickened overcrossing arc) of updates. Independent updates commute, as expressed by the $I3$ move of Figure~\ref{F:local_moves_machines} in Section~\ref{S:Machines}\ref{SS:Equivalence}.
 \item Local directionality (direction of the thickened arc) but no global `time line' within the machine.
 \item A flexible setup in which equivalent machines have different performance features.
 \end{enumerate}

\subsection{Summary of examples}

\begin{description}
\item[Classical information] The machines of Section~\ref{S:information}, whose arcs are coloured by entropies, represent distributed information processing. The difference between input and output entropies represents the capacity of the computation, and the computation is said to be \emph{optimal} if this number equals the mutual information of the input and the output. We exhibit three equivalent machines which represent computations with are locally optimal, locally suboptimal, and abstract. This paradigm, which we plan to study further in the future, takes the formalism of tangle machines into the realm of information theory.

\item[Adiabatic quantum computation]
Section~\ref{S:AQC} discusses how machines can represent networks of adiabatic quantum computations, and presents equivalent machines which perform the same computations, but with different energy gaps.

\item[Iteration]
By concatenating copies of the same machine one beside the other, machines may represent iterative computations and Markov chains. We exhibit three machines, one of which has a stochastic transition matrix, and two of which of which have a stochastic two-step transition matrix but not a one-step transition matrix. One of these represents a feed-forward system, while the other represents a feedback loop. This is discussed in Section~\ref{S:Recursion}.
\end{description}

\subsection{Acknowledgements}

The authors thank Louis Kauffman and Marius Buliga for useful comments. DM thanks also Dror Bar-Natan for useful suggestions.

\section{Scientific context}\label{SS:Context}


\subsection{Low dimensional topology to model computation}
The idea to model computations using tangle diagrams and related structures from low dimensional topology was pioneered by Louis Kauffman. Motivated by Spencer--Brown's \emph{Laws of Form} \citep{SpencerBrown:69}, Kauffman used knot and tangle diagrams to study automata \citep{Kauffman:94}, nonstandard set theory, and lambda calculus \citep{Kauffman:95,BuligaKauffman:13}. The diagrammatic calculus of braids (braids are a special class of tangles) also lies at the basis of topological quantum computing--- see \textit{e.g.} \citep{KauffmanLomonaco:04,Nayak:08}. 
Buliga has suggested to represent computations using a calculus of coloured tangles \citep{Buliga:11b}. 
In another direction, a different diagrammatic calculus, originating in higher category theory, has been used in the theory of quantum information--- see \textit{e.g.} \citep{AbramskyCoecke:09,BaezStay:11,Vicary:12}.

 We would argue that our approach is conceptually distinct from all previous approaches for the following reasons. First, on the practical level, we have not yet found direct overlap between our applications and the applications of other low-dimensional topological approaches to computation--- we do not know how we might describe \textit{e.g.} physical motion of anyons, nor do we know how previous diagrammatic approaches could naturally describe any of our examples. Secondly, tangle machines place primary emphasis on a distributive property of computation and of information fusion (compare~\citep{Roscoe:90}) via the R3 move of Figure~\ref{F:local_moves_machines1} in Section~\ref{F:Machine}\ref{SS:Equivalence}, as opposed to other approaches in which the lead role is played by associativity of a `stacking' operation. Thirdly, in contrast with most other approaches, tangle machines are coloured. Colours of registers represent information and are a fundamental part of our structure. Interactions are coloured by binary operations representing fusion or computation schemes, which may differ for different interactions. A fourth difference is that our interactions cannot be merged or split, so that we can graphically represent independence or indistinguishability of operations, which could not be deduced from the colours alone. A fifth difference is that only our overstrands are oriented and that their orientations are independent of one another. Thus there is no global time line and directionality is localized at crossings. For these reasons we do not believe that it is possible to usefully reformulate tangle machines in, for instance, the language of braided monoidal categories, that is the language of categorical quantum mechanics.


\subsection{Diagrams in the plane as brave new algebra}
The combinatorial paradigm of knot theory manifests a new philosophy of what constitutes \emph{algebra} \citep{Nelson:11}. For the combinatorial knot theorist, algebra no longer consists merely of formal manipulations of strings of symbols, but rather of local rewriting moves on labeled figures in the plane and in higher dimensions. This new philosophy of diagrammatic algebra has become particularly well established in the representation theory of quantum groups, in higher category theory, and in quantum field theory. Perhaps knot diagrams, tangle diagrams, and related objects are logical structures, algebraic structures, and categorical structures as much as they are topological structures (see \textit{e.g.} \citep{Kauffman:95}). In this paper, tangle machines are considered primarily as algebraic structures which capture and which highlight an underlying distributive aspect of computation and of information fusion. 

\subsection{An expanded notion of computation}

Turing machines are the heart of theory of computation and complexity theory \citep{Turing:37}. They formalize the notion of an algorithm or of an effective procedure, and they define the class of computable functions. There are profound interrelationships between Turing machines and low dimensional topology. For example, the classification of four-manifolds is undecidable due to the unsolvability of the word problem for finitely presented groups (see \textit{e.g.} \citep{Miller:92}). But there is an ongoing debate as to whether the non-mathematical Church--Turing thesis holds in general, namely, whether any (intuitively) computable function is realizable by a Turing Machine. It essentially questions the expressiveness of the Turing model in various non-(Turing) standard settings. To quote Copeland \citep{Copeland:04}:

\begin{quote}
 It is an open question whether there can be actual deterministic physical processes that, in the long run, elude simulation by a Turing machine, and in particular whether any such hypothetical process could usefully be harnessed in the form of a calculating machine (a hypercomputer) that could solve the halting problem for a Turing machine amongst other things. It is also an open question whether any such unknown physical processes are involved in the working of the human brain, and whether humans can solve the halting problem.
\end{quote}

The present paper suggests that coloured knots, tangles, spaces, and related structures can be computers. Indeed, the term \emph{tangle machine} imitates \emph{Turing machine}. The \emph{computation} of a tangle machine involves reading off colours of a chosen set of \emph{output registers} given a colouring of a chosen set of \emph{input registers} (assuming that the latter uniquely dictates the former). A tangle machine may thus capture a certain sort of network computation. In future work we plan to investigate tangle machine simulation of Turing machines and of neural nets. 

\section{Machines and machine equivalence}\label{S:Machines}

In this section we introduce the diagrammatic formalism of \emph{tangle machines}, which we call \emph{machines} for short.

\subsection{The set of labels of a machine: A quandle}\label{SS:Quandle}


We consider a set $Q$, whose elements we call \emph{colours}, equipped with a set $B$ of binary operations from $Q\times Q$ to $Q$. We think of elements of $Q$ as representing \emph{pieces of information} and of elements of $B$, which we call \emph{updates}, as representing \emph{information fusion} or \emph{basic computation}, although the precise interpretation of these terms is different in each of Sections~\ref{S:information},~\ref{S:AQC},~and~\ref{S:Recursion}.

Our updates are required to satisfy three properties:

\begin{description}
\item[Idempotence:] $x\trr x=x$ for all $x\in Q$ and for all $\trr\in B$.
\item[Reversibility:] The map $\trr y\colon\, Q\to Q$, which maps each colour $x\in Q$ to a corresponding colour $x\trr y\in Q$, is a bijection for all $(y,\trr)\in (Q,B)$. In particular, if $x\trr y = z\trr y$ for some $x,y,z\in Q$ and for some $\trr\in B$, then $x=z$. We interpret this condition to mean for example that information fusion does not forget information, because $x$ can uniquely be reconstructed from $x\trr y$ together with $\trr$ and $y$.
\item[Distributivity:] For all $x,y,z\in Q$ and for all $\trr,\brr \in B$:
\begin{equation}\label{E:Distributivity} (x\trr y)\brr z= (x\brr z)\trr (y\brr z)\enspace .\end{equation}
\noindent We interpret this equation to mean for example that information fusion eliminates redundancy. Thus, information $z$ which appeared once in $x\brr z$ and once in $y\brr z$ is not double-counted towards $(x\brr z)\trr (y\brr z)$.
\end{description}

We call $(Q,B)$ a \emph{$B$--family of quandles} or just a \emph{quandle}.

\begin{rem}
The low-dimensional topology literature contains several variants on our notion of a $B$--family of quandles. The usual definition of a quandle is the case when $B$ consists of only a single element $\trr$ and its inverse $\rrt$ (\textit{e.g.} \citep{Joyce:82}). Ishii \textit{et.al.} defined the notion of a \emph{$G$--family of quandles} \citep{Ishii:12}, in which elements of $B$ are indexed by a group $G$ and satisfy two additional compatibility relations. The case of an abelian group $G$ had been considered previously in \citep{Buliga:11a}. The set $B$ can indeed be turned into an abelian group in all of our examples, and the additional conditions are satisfied. But because we do not make use of this additional structure, we do not impose it. Our notion of a $B$--family of quandles follows Przytycki \citep{Przytycki:11} who named such a structure a \emph{multi-quandle}.
\end{rem}

We list several archetypal examples of $B$--families of quandles.

\begin{example}[Conjugation quandle]\label{E:QuantumGate}
Colours might be elements of a group $\Gamma$, and the operation might be conjugation:
\begin{equation}
x\brr y \ass y^{-1}xy\enspace .
\end{equation}
The pair $(\Gamma,\set{\brr})$ is called a \emph{conjugation quandle}. Such quandles feature in knot theory, \textit{e.g.} \citep{Joyce:82}.
\end{example}

\begin{example}[Linear quandle]\label{E:LinearQuandle}
Colours might be elements of a real vector space $Q$ and the operations might be convex combinations:
\begin{equation}
x\trr_s y \ass (1-s)x + sy \qquad s\in D\subseteq \mathds{R}\setminus\set{1}\enspace.
\end{equation}
The pair $\left(Q,\set{\trr_s}_{s\in D}\right)$ is called a \emph{linear quandle}. Our examples in Sections~\ref{S:information},~\ref{S:AQC},~and~\ref{S:Recursion} all involve linear quandles. 
\end{example}


\begin{example}[Loglinear quandle]\label{E:LogLinear}
In the same setting as Example~\ref{E:LinearQuandle}, consider the operations:
\begin{equation}
x\,\bar{\trr}_s\, y \ass x^{1-s}y^{s} \qquad s\in D\subseteq \mathds{R}\setminus\set{1}\enspace .
\end{equation}
The pair $\left(Q,\set{\bar{\trr}_s}_{s\in D}\right)$ is called a \emph{loglinear quandle}. In \citep{CarmiMoskovich:14c} we exhibited several standard information fusion operations as quandle operations of quotients of loglinear quandles.
\end{example}

\subsection{Recursive definition of tangle machines}\label{SS:TangleMachines}

The fundamental building block of a machine is an \emph{interaction}. The simplest interaction is graphically depicted as
\begin{equation}
\psfrag{u}[r]{\small $x$}
\psfrag{y}[c]{\small $y$}
\psfrag{x}[l]{\small $x \trr y$}
\includegraphics[width=0.1\textwidth]{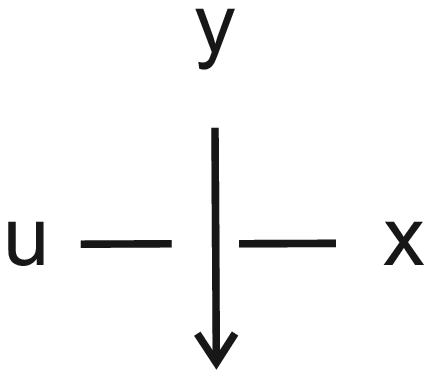}
\end{equation}
This interaction describes initial information $x$ (called the \emph{input patient}) being updated by new information $y$ (called the \emph{agent}) to obtain updated information $x\trr y$ (called the \emph{output patient}). The updating operation $\trr$ may differ for different interactions. 
The colours $x$, $y$, and $x\trr y$ are elements of a quandle $(Q,B)$. We name the strands being coloured as \emph{registers}. The assignments of colours to registers and of binary operations to interactions is called \emph{colouring}.

The agent in an interaction may update multiple registers. In this case the agent is drawn as a thick line. For example (with colours suppressed):

\begin{equation}
\includegraphics[width=0.8in]{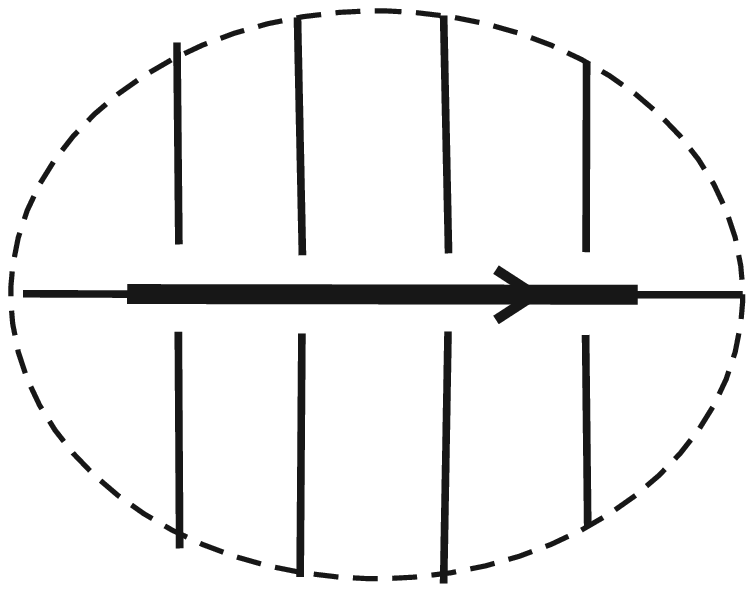}
\end{equation}

A general tangle machine is obtained by concatenating a disjoint union of a finite number of interactions and building blocks which look like\ \ $\includegraphics[width=20pt]{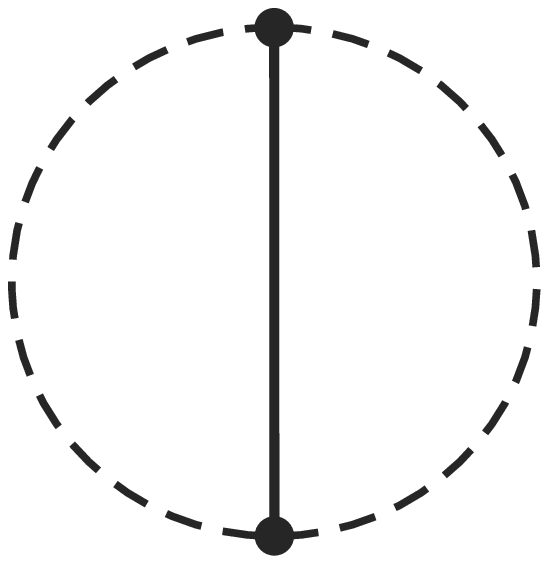}$. Concatenation is the process of connecting endpoints of a tangle machine.

\begin{equation}
\includegraphics[width=0.9\linewidth]{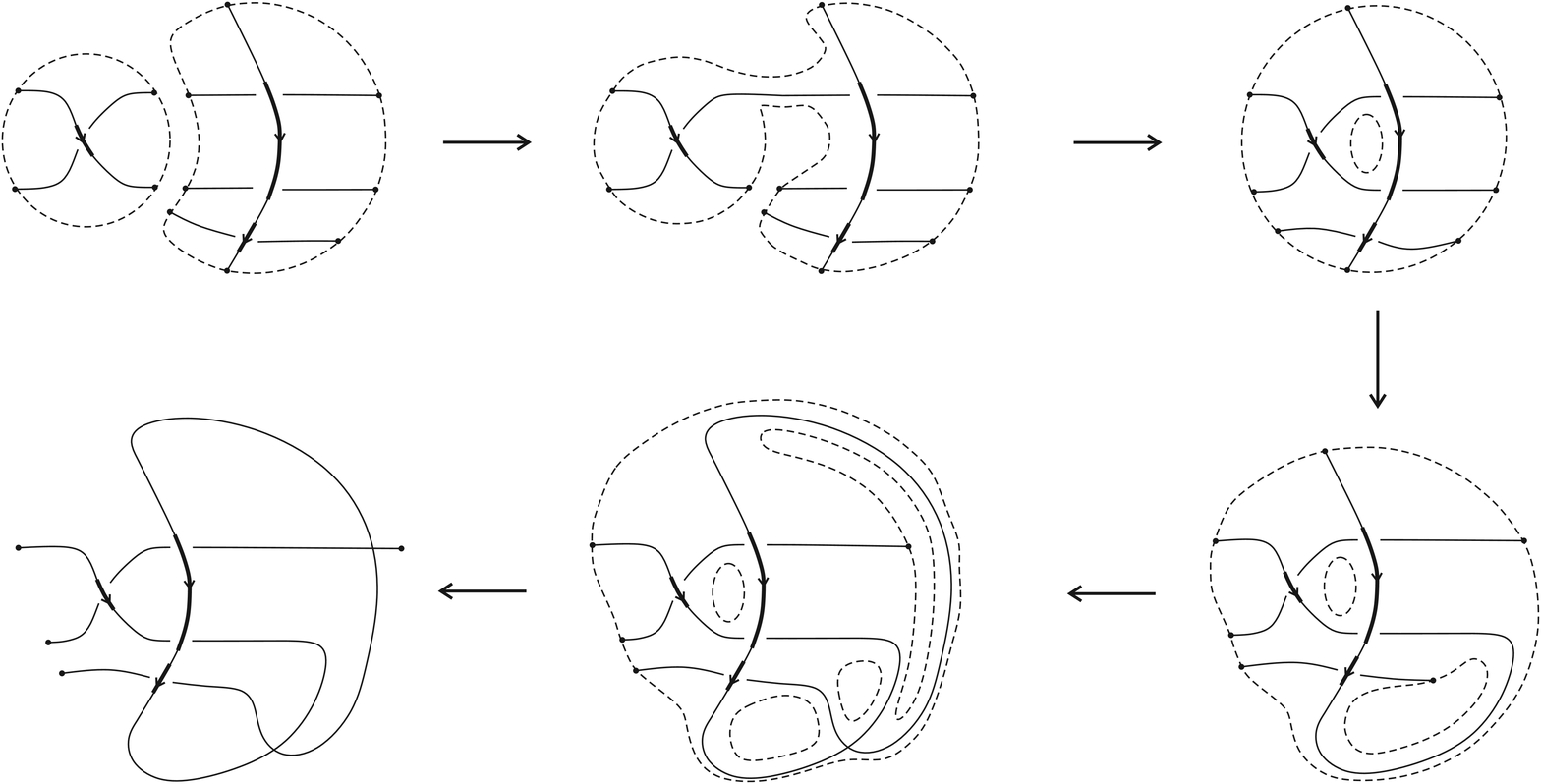}
\end{equation}

The dotted line around the interaction, called the \emph{firmament}, and the endpoints of the interaction, should be thought of as living ``at infinity''. The purpose of the firmament is to facilitate the definition of concatenation. As such, the final stage in our construction is to erase the firmament and to draw rays from each tangle endpoint to infinity which intersect the tangle diagram only transversely at double-points (if at all). The choice of these `rays to infinity' is essentially arbitrary. We do not distinguish between tangle machines which differ only in the choice of these rays. 

\begin{rem}
Our diagrammatic model of machines as concatenated interactions is inspired by diagrammatic formalisms in low dimensional topology. \emph{Combinatorial~knot~theory} studies knots as planar diagrams instead of as embedded objects in $3$--space. These diagrams may be decomposed into \emph{tangles} \citep{Conway:70}. Knots and tangles are modified by \emph{local moves}, which replace one tangle within a knot by another. Knots are thus revealed to be algebraic objects arising as concatenations of crossings (which are very simple tangles) in the plane \citep{Jones:99}. Dropping the requirement that concatenation be planar, Kauffman defined \emph{virtual tangles} \citep{Kauffman:99}. A strengthening of the equivalence relation imposed on virtual tangles gives rise to w-tangles. Our diagrammatic construction is most similar to the diagrammatic calculus of w-tangles \citep{BarNatanDancso:13}, which form an algebra over a \emph{modular~operad} \citep{GetzlerKapranov:98}. The differences are that our diagrams are coloured, that we allow multiple quandle operations, and that our interactions cannot be split or merged. Also, as in the theory of \emph{disoriented~tangles}, no compatibility condition is imposed for directions of concatenated agents \citep{ClarkMorrisonWalker:09}.
\end{rem}

The rigourous definition of a tangle machine is deferred to the appendix. Examples of tangle machines are given below and are scattered throughout the paper.

\begin{equation} 
\begin{minipage}[t]{0.4\linewidth}
\psfrag{x}[c]{\small $y$}
\psfrag{y}[c]{\small $x$}
    \includegraphics[width=0.95\textwidth]{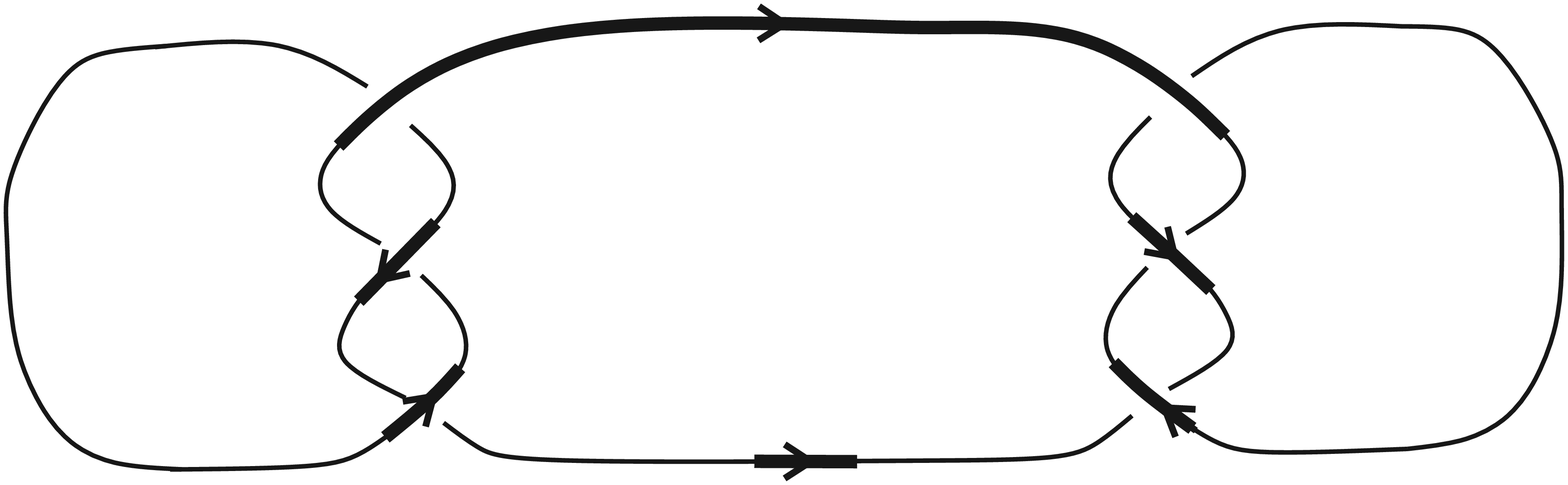}
\end{minipage}
 \hspace{7pt}
\begin{minipage}[t]{0.43\linewidth}
\centering
\psfrag{x}[c]{\small $y$}
\psfrag{y}[c]{\small $x$}
\psfrag{a}[c]{\small $x$}
\psfrag{b}[c]{\small $y$}
\psfrag{c}[c]{\small $v$}
\psfrag{d}[r]{\small $w$}
    \includegraphics[width=0.5\textwidth]{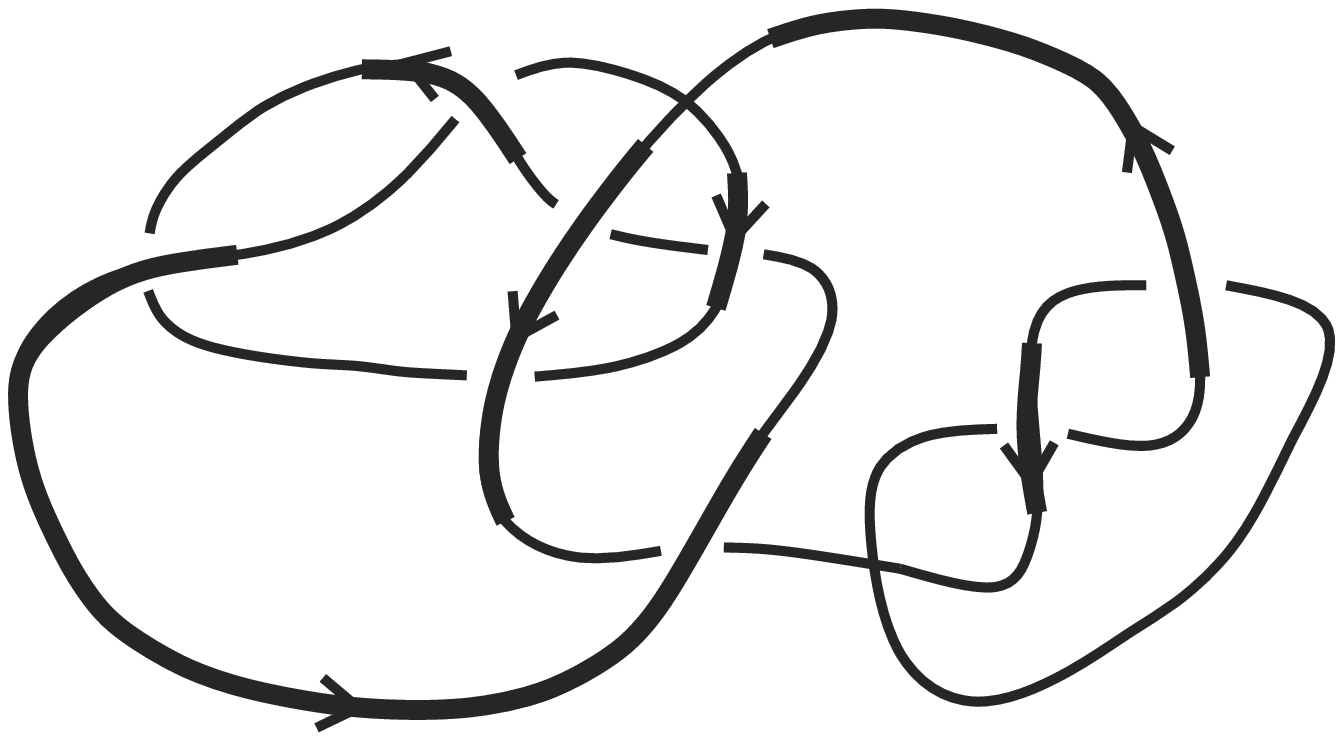}
\end{minipage}
\end{equation}

\subsection{Machine equivalence}\label{SS:Equivalence}

The main feature of machines is their natural local notion of equivalence, described here and assembled into a concise definition in the appendix.

\begin{figure}[htb]
\centering
\psfrag{a}[c]{\small $\trr$}
\psfrag{b}[c]{\small $\rrt$}
\psfrag{V}[c]{\small \emph{I1}}
\psfrag{x}[c]{\small $x$}
\psfrag{T}[c]{\small \emph{VR1}}\psfrag{R}[c]{\small \emph{VR2}}\psfrag{S}[c]{\small \emph{VR3}}
\psfrag{Q}[c]{\small \emph{SV}}\psfrag{D}[c]{\small \emph{I2}}\psfrag{E}[c]{\small \emph{FM1}}\psfrag{F}[c]{\small \emph{FM2}}\psfrag{C}[c]{\small \emph{I3}}\psfrag{Y}[c]{\small \emph{ST}}
\psfrag{X}[c]{\small \emph{ST}}
\includegraphics[width=0.85\textwidth]{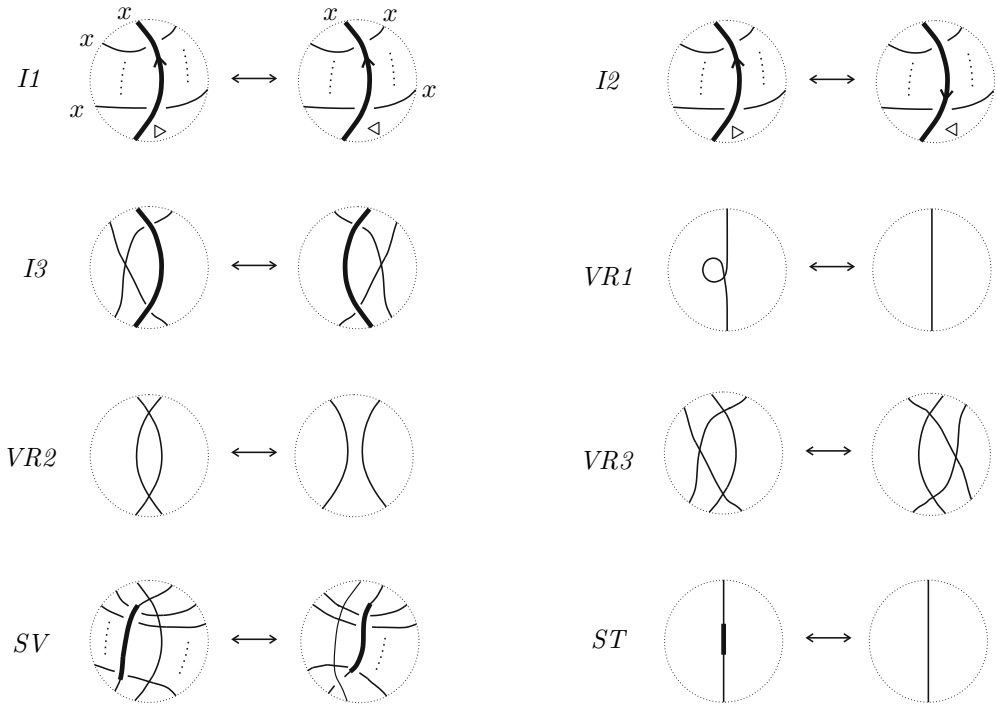}
\caption{\label{F:local_moves_machines} \small Cosmetic moves for machines. Where directions are not indicated, the meaning is that the move is valid for any directions, and the same for colourings.}
\end{figure}

\begin{figure}[htb]
\centering
\psfrag{T}[c]{\small \emph{VR1}}\psfrag{R}[c]{\small \emph{VR2}}\psfrag{S}[c]{\small \emph{VR3}}
\psfrag{Q}[c]{\small \emph{SV}}\psfrag{D}[c]{\small \emph{R1}}\psfrag{A}[c]{\small \emph{R2}}\psfrag{B}[c]{\small \emph{R3}}\psfrag{C}[c]{\small \emph{UC}}
\psfrag{X}[c]{\small \emph{ST}}
\includegraphics[width=0.85\textwidth]{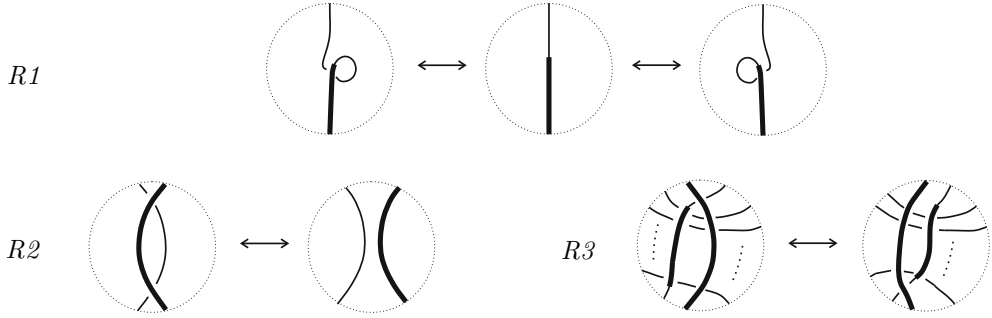}
\caption{\label{F:local_moves_machines1} \small Reidemeister moves for machines, valid for any directions of the agents and for any colouring.}
\end{figure}

First, we do not ascribe physical meaning to colours, but only to differences between colours. Thus, if change the colouring of a machine $M$ by an action of an automorphism of $(Q,B)$ inside a disc $D$, where $M$ does not intersect the boundary of $D$, then the resulting machine $M^\prime$ is considered to be equivalent to $M$.

Secondly, as in graph theory, intersections between edges in diagrams of machines `do not really exist', and can be added or taken away at will by one of the modifications $V\!R1$, $V\!R2$ and $V\!R3$ in Figure~\ref{F:local_moves_machines}. This amounts to choosing different concatenating lines when recursively building the machine out of interactions. Moves $I1$, $I2$, and $I3$ relate local pictures which express the same inputs changing to the same outputs as a result of the same agent. And move $ST$ allows us to add and delete agents which do not act on anything.

\begin{figure}\label{E:disk}
\centering
\includegraphics[width=0.95\linewidth]{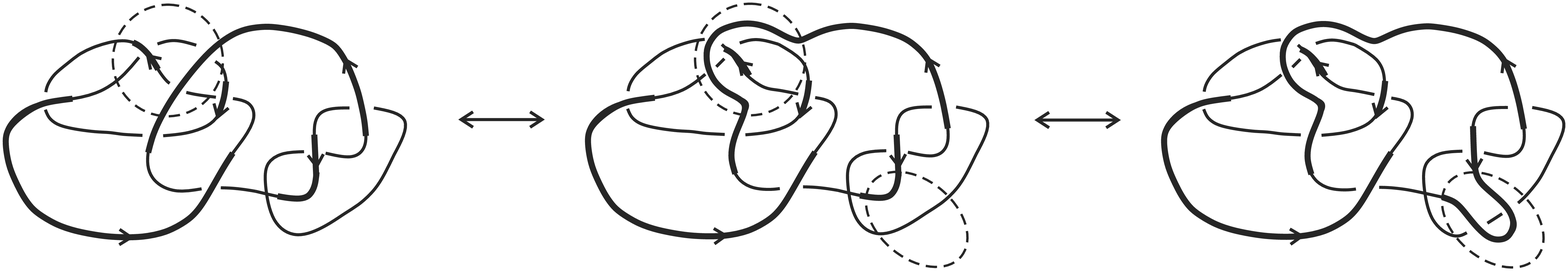}
\caption{\label{F:disk}An example of machine equivalence.}
\end{figure}

Third, updates performed by a single agent should be thought of as simultaneous. Thus, the two diagrams below, whose diagrams differ by permutation of input-output pairs (on the LHS the agent, indicated by the thick line, appears first to update process $A$ and then process $B$, while on the RHS it appears first to update process $B$ and then process $A$), depict equivalent machine:

\begin{equation}\label{E:r3a3}
\begin{minipage}{2.2in}
\centering
\psfrag{S}{}\psfrag{a}[c]{$A$}\psfrag{b}[c]{$B$}
\includegraphics[width=1.9in]{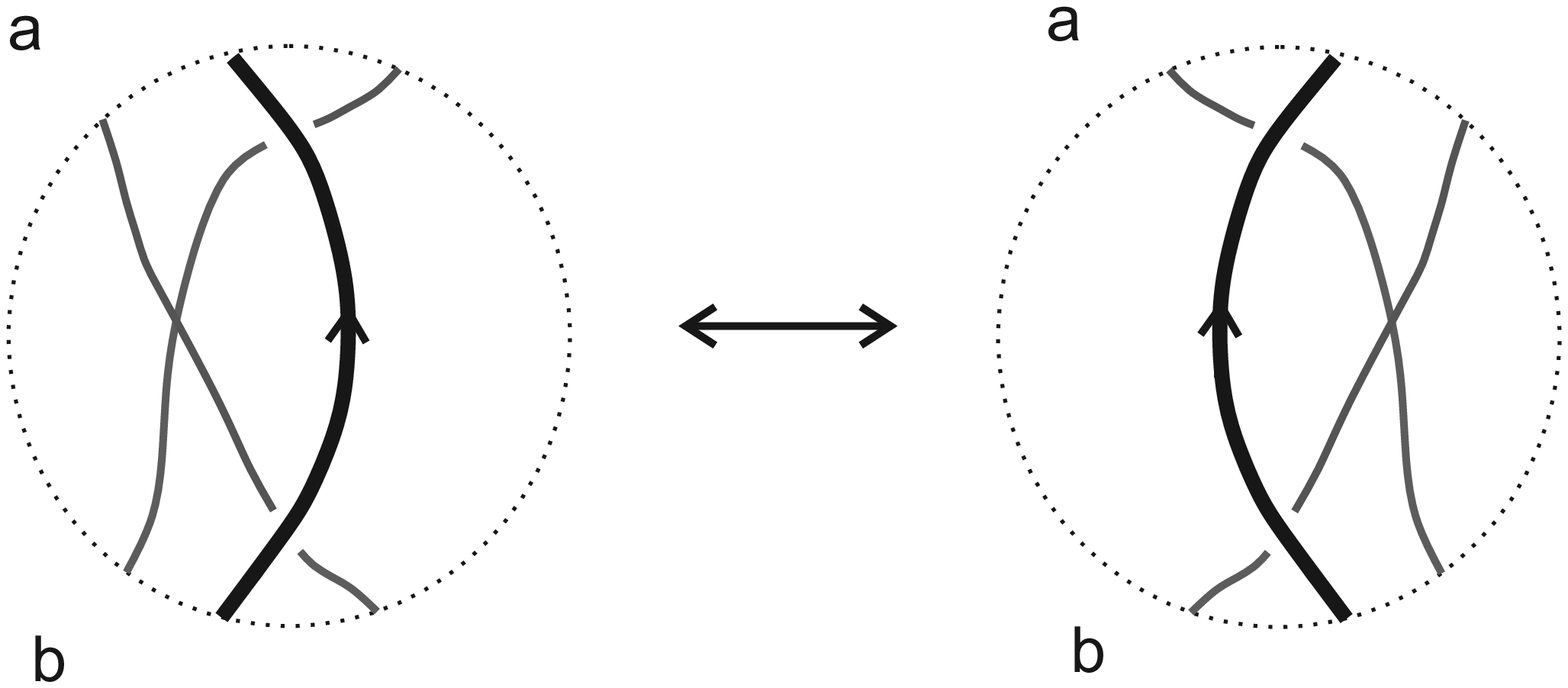}
\end{minipage}
\end{equation}

Fourth, the Reidemeister moves, R1, R2 and R3 of Figure~\ref{F:local_moves_machines1} embody the defining axioms of $(Q,B)$. This is illustrated in \eqref{eq:r1}, \eqref{eq:r2}, and \eqref{eq:r3}, which reflect idempotence, reversibility, and distibutivity respectively (in each equation designated colours on either side of the arrow are equal). Note that reversibility implicitly defines an inverse operation $\rrt$ for each $\trr\in B$ such that $(x\trr y)\rrt y= x$ for all $x,y\in Q$. Machines related via a finite sequence of Reidemeister moves are considered equivalent.

\begin{subequations}
\begin{equation}
\label{eq:r1}
\psfrag{a}[c]{\small $\overline{x}$}
\psfrag{b}[c]{\small $x$}
\psfrag{c}[c]{\small $\overline{{x \trr x}}$}
\includegraphics[width=0.38\textwidth]{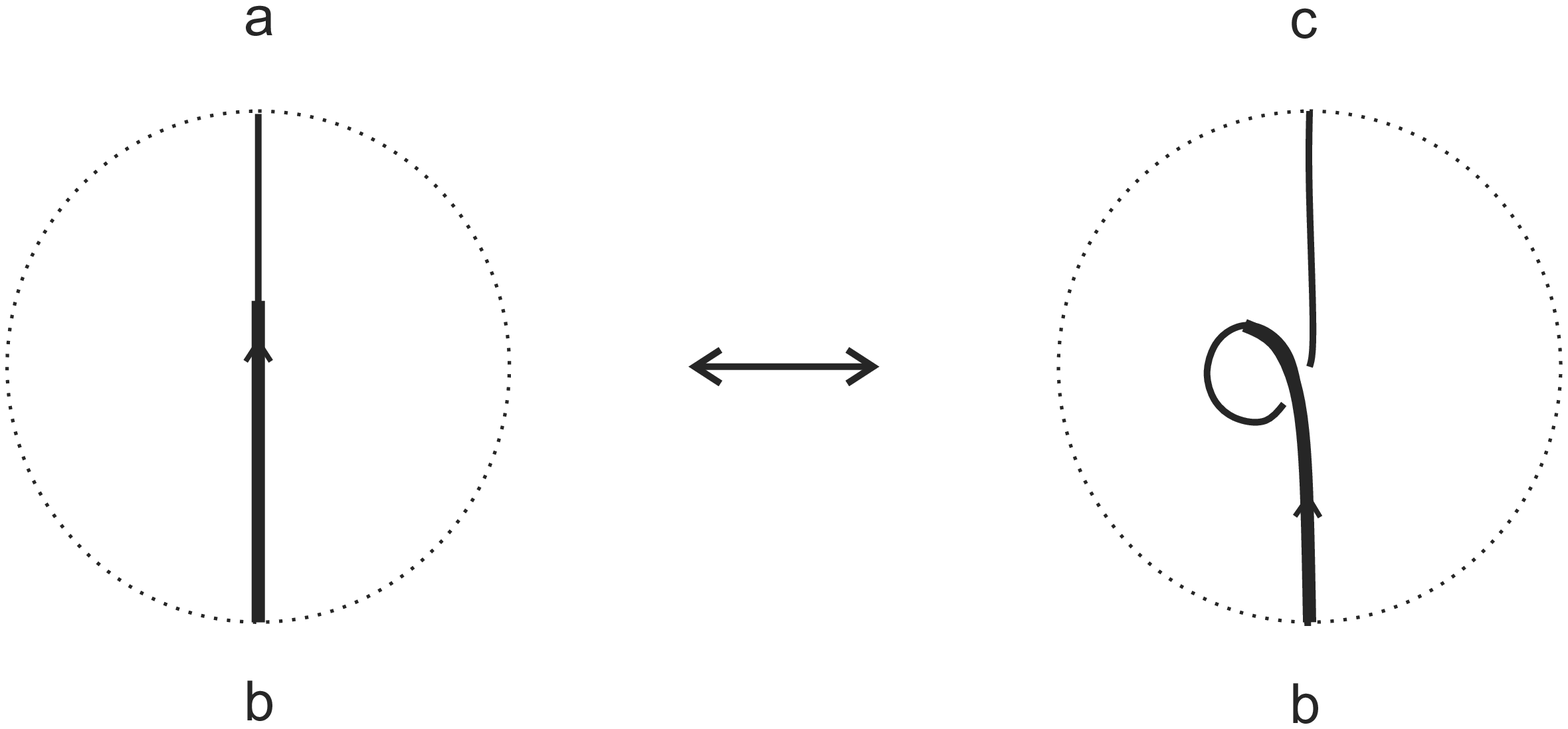}
\end{equation}
\vspace{3pt}
\begin{equation}
\label{eq:r2}
\psfrag{a}[c]{\small $y$}
\psfrag{c}[c]{\small $y$}
\psfrag{e}[c]{\small $y$}
\psfrag{h}[c]{\small $y$}
\psfrag{b}[c]{\small $x$}
\psfrag{d}[c]{\small $\overline{x}$}
\psfrag{f}[c]{\small $x$}
\psfrag{I}[l]{\small $\overline{{(x \trr y) \rrt y}}$}
\includegraphics[width=0.38\textwidth]{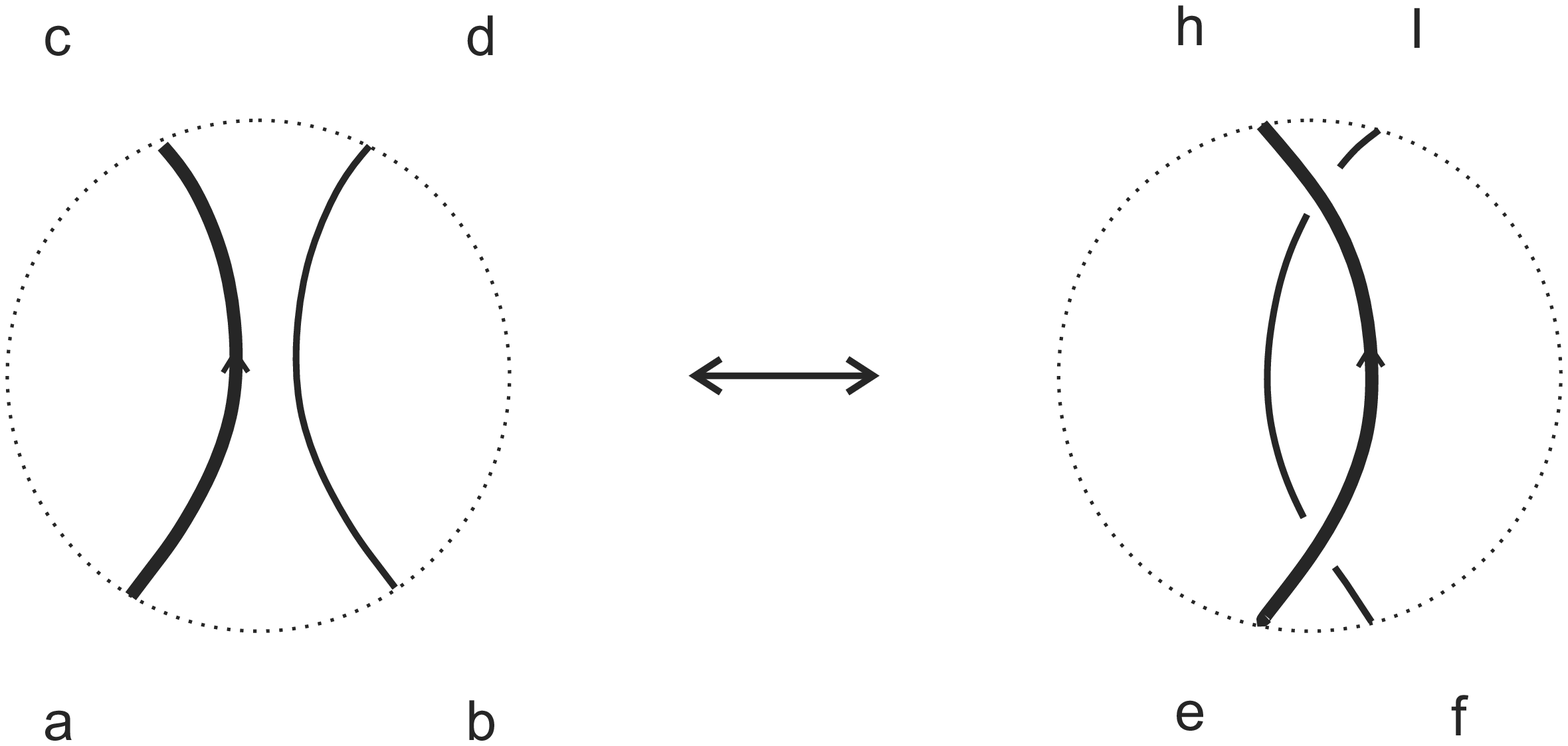}
\end{equation}
\vspace{3pt}
\begin{equation}
\label{eq:r3}
\psfrag{a}[c]{\small $y$}
\psfrag{c}[c]{\small $z$}
\psfrag{z}[c]{\small $x$}
\psfrag{a}[c]{\small $x$}
\psfrag{b}[c]{\small $y$}
\psfrag{e}[l]{\small $y \brr z$}
\psfrag{d}[l]{\small $\overline{(x \trr y) \brr z}$}
\psfrag{f}[l]{\small $\overline{(x \brr z) \trr (y \brr z)}$}
\includegraphics[width=0.38\textwidth]{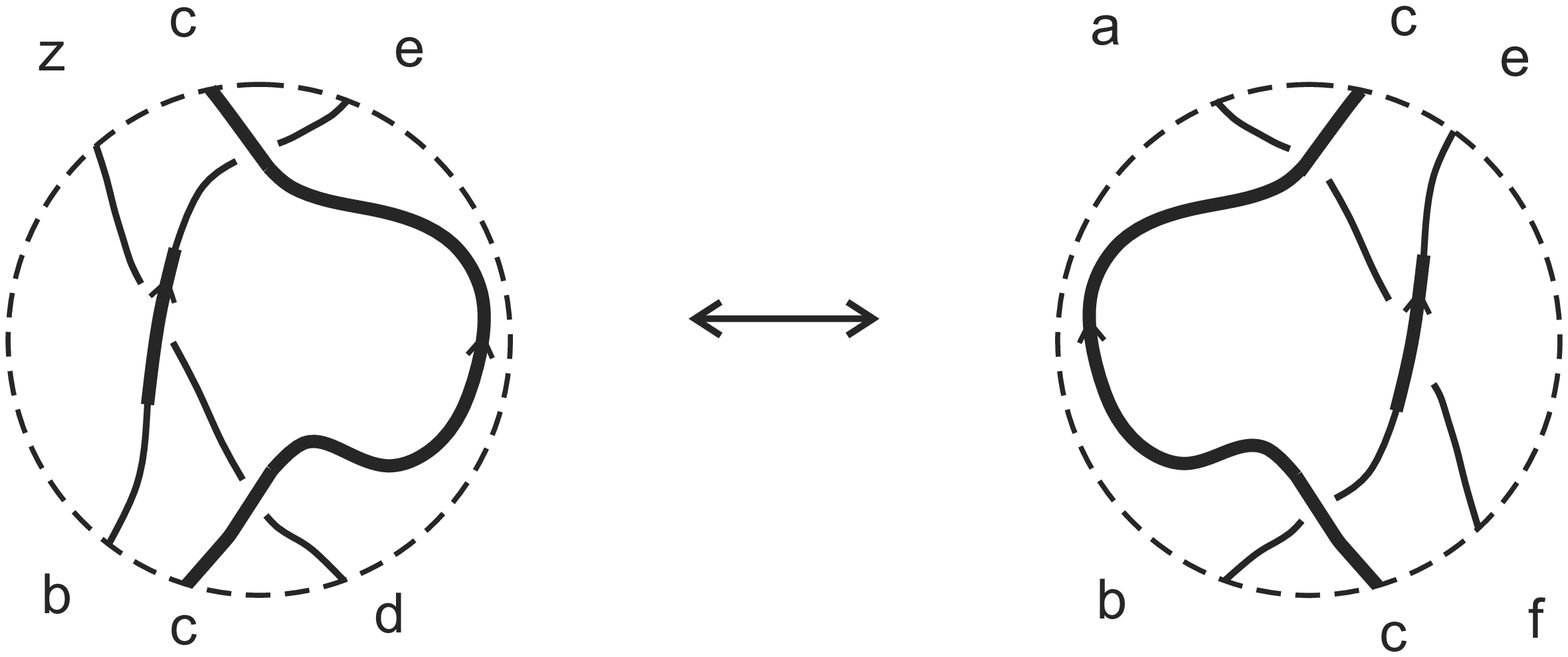}
\end{equation}
\end{subequations}

\section{Machines and information}\label{S:information}

The concept of computation is broad, and extends beyond calculating the answer to a prescribed problem.
Perhaps the most general characterization of computation is that it is `a manipulation or processing of information'. Computation and information are intertwined, and these two concepts rely heavily on one another. 

In this section, machines are conceived of as a class of networks for distributed information processing. The colours represent information entropies. The information processing capacity associated with an interaction, called its \emph{local capacity}, is  defined to be a difference between initial and terminal colours. A machine $M$ represents a network within which information is processed and sent further down to other interactions or registers. A machine equivalent to $M$ has the same information processing capacity as $M$, but its local capacities may be different.

Our definitions in this section follow \citep{CoverJoy:06}. We keep our discussion as informal as possible.


\subsection{Preliminary definitions}\label{SS:infoprelim}

An \emph{information channel} is an apparatus through which messages are transmitted from one location to another. In practical situations, a message entering the channel on one end will emerge corrupted on the other end. It is convenient to think of a message as a sequence of zeroes and ones. An information channel is characterized by its \emph{capacity}, that is the maximal rate at which messages may be transmitted with a `negligible' loss of information.
Entropy is a measure of information, or rather, of uncertainty. If a message is constructed by sampling $N$ independent identically distributed (iid) binary random variables, then Shannon's Source Coding Theorem \citep{Shannon:48} tells us that, for \emph{typical sequences}, the entropy times $N$ is nearly the number of information units (\textit{e.g.} bits) required to encode a message so that it can reliably be recovered by a receiver.

\emph{Compressible messages} exhibit some kind of pattern ($H<1$), and these admit shorter descriptions than the length of the message itself. This is the key principle underlying message compression. \emph{Incompressible} messages are messages for which randomness inhibits descriptions shorter than the message own length (\textit{i.e.} $H = 1$).

A general computing device (\emph{e.g.} a universal Turing machine) requires two distinct inputs. The first input $\mathcal{X}_0$ is a stream of data that is read and manipulated by the machine according to instructions given by the second input $\mathcal{X}_1$. Both inputs $\mathcal{X}_0$ and $\mathcal{X}_1$ and the \emph{result} of a computation $\mathcal{X}_{\text{out}}$ are assumed to be typical binary sequences.

\subsection{Information processing by machines}\label{SS:infomachines}

A machine describing an information processing network is a concatenation of interactions. Each of its registers is coloured by a real number representing an entropy. The colour of an agent register represents the entropy of a programme typical sequence, while colours of input registers represent entropies of data typical sequences. The agent register is equipped with a parameter $s \in (0,1)$ which may represent some (input-independent) property of the computing device itself. The colour of the output corresponding to input $H(\mathcal{X}_0)$ is:

\begin{equation}
\label{eq:fusion}
H(\mathcal{X}_0) \trr_s H(\mathcal{X}_1) \ass  (1-s) H(\mathcal{X}_0) + s H(\mathcal{X}_1)\enspace .
\end{equation}

If $H(\mathcal{X}_0) > H(\mathcal{X}_1)$ then the output entropy is strictly lower than the input entropy, \textit{i.e.} $H(\mathcal{X}_0) \trr_s H(\mathcal{X}_1) < H(\mathcal{X}_0)$.

\begin{figure}[htb]
\centering
\psfrag{a}[r]{\small $_{H(\mathcal{X}_0)}$}
\psfrag{b}[c]{\small $_{H(\mathcal{X}_1)}$}
\psfrag{c}[l]{\small $_{H(\mathcal{X}_0) \trr_s H(\mathcal{X}_1)}$}
\psfrag{a1}[r]{\small $\mathcal{X}_0$}
\psfrag{b1}[c]{\small $\mathcal{X}_1$}
\psfrag{c1}[l]{\small $\mathcal{X}_{\text{out}}$}
\psfrag{s}[c]{\small $s$}
\includegraphics[width=0.65\textwidth]{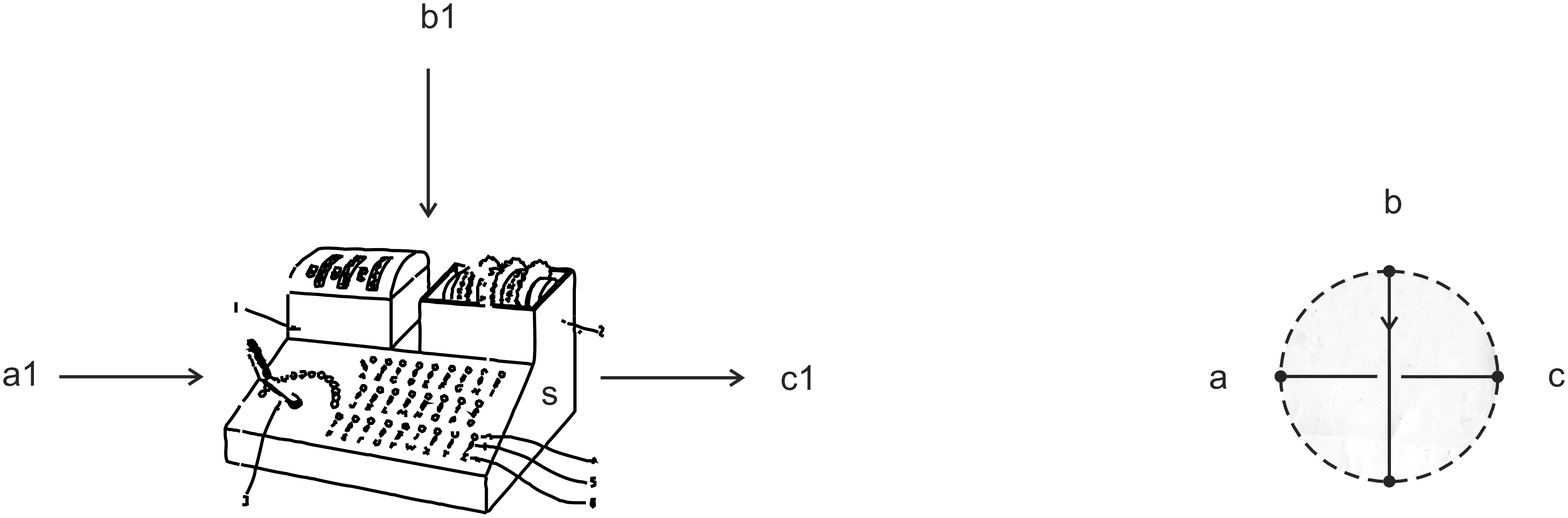}
\caption{\label{F:ComputerCrossing} The computation and the corresponding interaction between entropies.}
\end{figure}

Thus, the machine computes $\mathcal{X}_{\text{out}}$ by applying the instruction data steam $\mathcal{X}_1$ to the input data stream $\mathcal{X}_0$, and the entropy of $\mathcal{X}_{\text{out}}$ is $H(\mathcal{X}_0) \trr_s H(\mathcal{X}_1)$. See Figure~\ref{F:ComputerCrossing}.

\subsection{Capacity}\label{SS:capacity}

In this section we describe various capacities associated to machines, which provide a measure of how `good' a computation is. Our analysis of a computing device whose internal workings are unknown to us focusses on discrepancies between its input and output streams. Suppose that we wish to know if the computation is meaningful in some sense. If no additional restrictions are made, then ``meaningful'' might mean that computations produce intelligible answers which could read off by a human operator. Translating this requirement into the language of preceding paragraphs, the output stream is expected to appear `less random' than the input stream. According to this paradigm, computation and compression are literally the same thing. A `good computation' is one which compresses $\mathcal{X}_0$ as much as possible, \emph{given $\mathcal{X}_1$}. In the language of information theory, the optimal output $\mathcal{X}_{\text{out}}$ has entropy equal to the \emph{conditional entropy} $H(\mathcal{X}_1 \mid \mathcal{X}_0)$. The \emph{channel capacity} of the computing device is defined as the \emph{mutual information}:
\begin{equation}
\label{eq:mui}
I(\mathcal{X}_1 : \mathcal{X}_0) \ass H(\mathcal{X}_1) - H(\mathcal{X}_1 \mid \mathcal{X}_0)\enspace.
\end{equation}

The \emph{capacity} of a \emph{process} (that is, a chain of registers connected by concatenation and by being input-output pairs of an interaction) is the entropy of its initial register minus the entropy of its terminal register. For example, for an interaction with a single input-output pair:
\begin{equation}
\label{eq:cc}
\mathrm{Cap}_s\left(\;\ \raisebox{5pt}{\ccross{_{\mathrm{In}}}{}{_{\mathrm{Out}}}} \; \; \;\   \right) \ass\ \underbrace{H(\mathcal{X}_0)}_{\mathrm{In}} - \underbrace{H(\mathcal{X}_0) \trr_s H(\mathcal{X}_1)}_{\mathrm{Out}}\enspace .
\end{equation}
An interaction is \emph{optimal} if its capacity equals its mutual information:
\begin{equation}
H(\mathcal{X}_0) - H(\mathcal{X}_0) \trr_s H(\mathcal{X}_1) = I(\mathcal{X}_0 : \mathcal{X}_1)\enspace ,
\end{equation}
\noindent which occurs when $H(\mathcal{X}_0) \trr_s H(\mathcal{X}_1) =H(\mathcal{X}_0 \mid \mathcal{X}_1)$.

The \emph{global capacity} of a machine is the set of all capacities of its processes.

\subsection{Equivalent machines}\label{SS:infoequiv}

\begin{figure}[htb]
\centering
\psfrag{a}[c]{\small $_{H(0)}$}
\psfrag{b}[c]{\small $_{H(1)}$}
\psfrag{c}[c]{\small $_{H(2)}$}
\psfrag{d}[c]{\small $\; _{H(0) \trr_t H(2)}$}
\psfrag{e}[l]{\small ${}_{H(1 \mid 2)}$}
\psfrag{e1}[c]{\small ${}_{H(1 \trr 0)}$}
\psfrag{f}[r]{\small $_{H(1 \mid 0, 2)}$}
\psfrag{g}[c]{\small ${}_{G}$}
\psfrag{v}[c]{\small $\trr_s$}
\psfrag{w}[c]{\small $\trr_t$}
\psfrag{r}[c]{\small \emph{abstract}}
\psfrag{s}[c]{\small \emph{locally suboptimal}}
\psfrag{t}[c]{\small \emph{locally optimal}}
\includegraphics[width=0.85\textwidth]{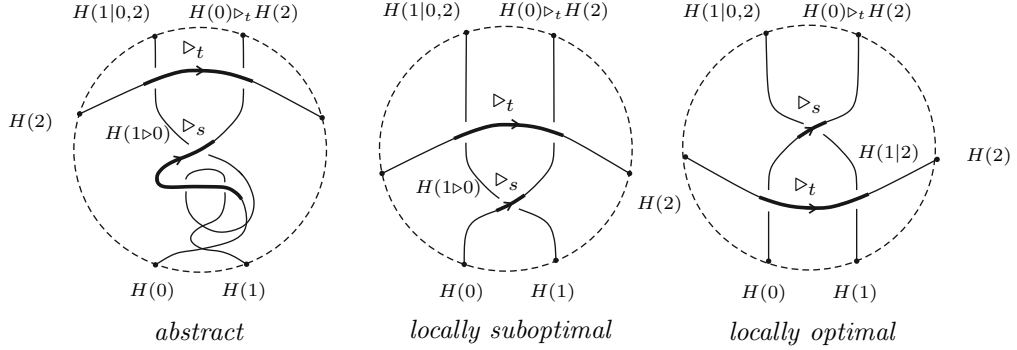}
\caption{\label{fig:procmachines}Equivalent machines with the same global information processing capacities. The middle and right machines are feasible whereas the left machine is abstract. While all of them are globally optimal only the rightmost machine is also locally optimal.}
\end{figure}

Consider the three equivalent machines in Figure~\ref{fig:procmachines}.
As the three machines are equivalent, they have the same global capacities. But the capacities of their interactions are different, and the leftmost machine represents an impossible, \emph{abstract} computation.

Set the following values of $t$ and $s$:
\begin{equation}
\label{eq:ts}
t \ass \frac{H(1)-H(1 \mid 2)}{H(1)-H(2)}, \quad s \ass \frac{H(1 \mid 2) - H(1 \mid 0,2)}{H(1 \mid 2) - H(0) \trr_t H(2)}\enspace .
\end{equation}

In order to assure that $t,s\in(0,1)$, we choose our entropies so that:

\begin{equation}
H(1 \mid 2) > H(2), \; \; \; H(1 \mid 0,2) > H(0) \trr_t H(2)\enspace,
\end{equation}
\noindent which essentially describe the extent to which the sources, $\mathcal{X}_0$, $\mathcal{X}_1$, and $\mathcal{X}_2$, are statistically dependent. This is illustrated by the following Venn diagrams:
\begin{equation}
\psfrag{a}[c]{\small $H(1 \mid 2)$}
\psfrag{b}[c]{\small $H(1 \mid 0,2)$}
\psfrag{y}[c]{\small $\mathcal{X}_2$}
\psfrag{x}[c]{\small $\mathcal{X}_1$}
\psfrag{z}[c]{\small $\mathcal{X}_0$}
\includegraphics[width=0.60\textwidth]{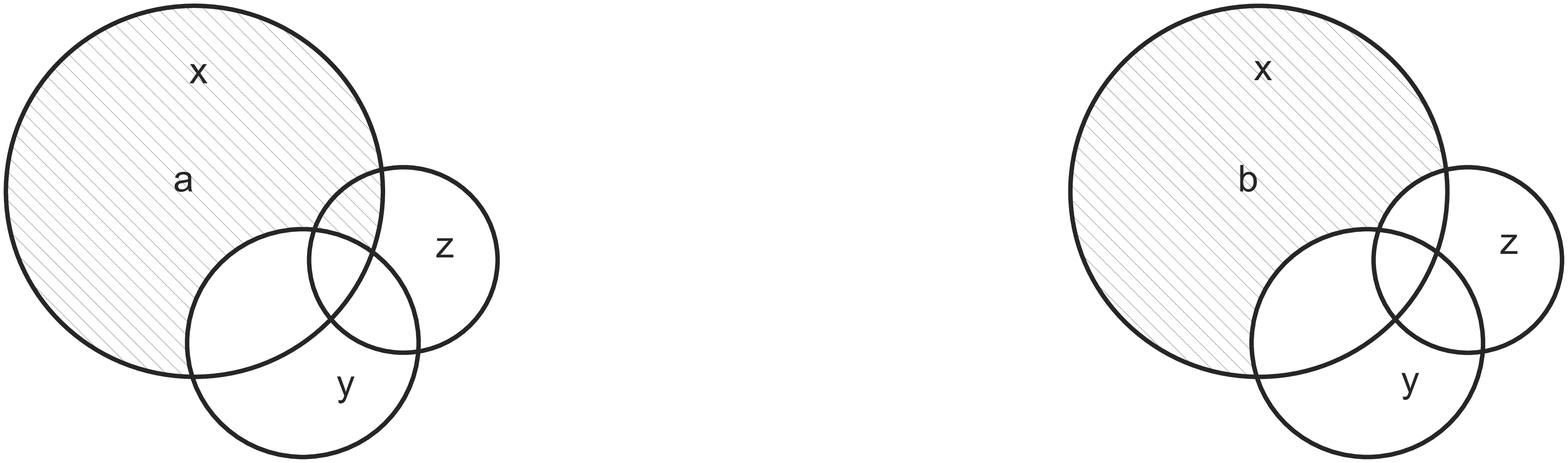}
\end{equation}

\begin{figure}[htb]
\centering
\psfrag{a}[c]{\small $_{H(0)}$}
\psfrag{b}[c]{\small $_{H(1)}$}
\psfrag{c}[c]{\small $_{H(2)}$}
\psfrag{x}[c]{\small $\mathcal{X}_1$}
\psfrag{y}[c]{\small $\mathcal{X}_2$}
\psfrag{z}[c]{\small $\mathcal{X}_0$}
\psfrag{d}[c]{\small $\; _{H(0) \trr_t H(2)}$}
\psfrag{e}[c]{\small \emph{process capacity in} $I(M)$}
\psfrag{f}[r]{\small $_{H(1 \mid 0, 2)}$}
\psfrag{g}[c]{\small \emph{local capacities at crossings} $\mathrm{Cap}$}
\psfrag{v}[c]{\small $\trr_s$}
\psfrag{w}[c]{\small $\trr_t$}
\psfrag{n}[c]{\small $I(1 : 2)$}
\psfrag{m}[c]{\small $I(1 : 2 \mid 0)$}
\psfrag{o}[c]{\small $I(1 : 2, 0)$}
\includegraphics[width=0.85\textwidth]{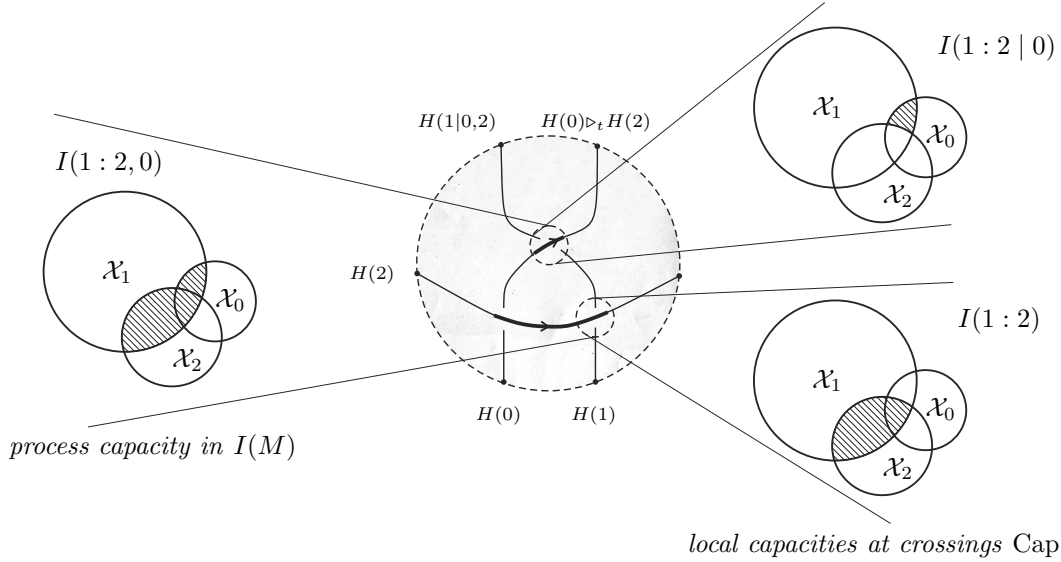}

\caption{\small Optimal information processing in the rightmost (locally optimal) machine in {Figure~\ref{fig:procmachines}}.}
\label{fig:capacities}
\end{figure}

All three machines are globally optimal, but the local capacities for the three machines in Figure~\ref{fig:procmachines} are different. In the rightmost machine, by our choices of $t$ and $s$, each interaction is locally optimal--- see Figure~\ref{fig:capacities}. This is no longer true for the middle machine, which has a register labeled $H(1 \trr 0) \ass H(1) \trr_s H(0)$, which may not equal $H(1 \mid 0)$. In this case, the middle machine contains a non-optimal interaction. The left machine involves the inverse operation $\rrt_s$, so that its colour $H(1) \triangleleft_s H(0)$ might be negative. The idea of negative entropies may sound absurd, but nevertheless the leftmost machine in Figure~\ref{fig:procmachines} is equivalent to a machine all of whose computations are feasible, and in fact even optimal. In view of this, we may think of this machine as a sort of \emph{abstract} information processing scheme.

\section{Adiabatic quantum machines}\label{S:AQC}

Some paradigms for quantum computation do away with the conventional circuit model. Adiabatic quantum computation is one such approach \citep{Farhi:2000}. The idea behind it rests on the Adiabatic Theorem in Quantum Mechanics which roughly states that a (quantum) system remains in its ground state when subjected to environmental perturbations, as long as these act slowly enough and as long as there is a gap between the ground state and the rest of the Hamiltonian's spectrum. Adiabatic quantum computation makes use of this fact by adiabatically evolving a simple Hamiltonian $H_0$, which can be thought of as a problem whose solution (the ground state) is easy, into a different and perhaps more complicated Hamiltonian $H_1$ whose ground state is the solution to the problem at hand. The computation initializes the system in its ground state, the ground state of $H_0$, and then slowly evolves its Hamiltonian to $H_1$. This process is called \emph{quantum annealing}. By the Adiabatic Theorem, the system remains in its ground state throughout the evolution process, and the computation concludes at the ground state of $H_1$, that is the sought-after solution.

The computational difficulty of this procedure is inversely proportional to the square of the minimal energy gap between the ground state and the rest of the spectrum, namely to the square of $g \ass \lambda_1 - \lambda_0$, where $\lambda_{i+1} \geq \lambda_i$ are the underlying energy eigenvalues of the Hamiltonian. 

We introduce an adiabatic quantum machine (AQC). Strictly speaking, this is a one-parameter family of tangle machines. For $s\in (0,1)$, consider the quandle $Q_s$ whose elements are self-adjoint operators over a Hilbert space of dimension $2^N$ and whose operation is $x \trr_s y \ass (1-s)x + s y$. In most cases $N$ stands for a number of qubits, and $N$ is always fixed. As $s$ evolves from $0$ to $1$, a machine $M_0$ coloured by a trivial quandle $Q_0$ evolves through machines $M_s$ coloured by $Q_s$. The algebraic structure $\lim_{t\to 1}Q_s$ is not a quandle (reversibility fails) but the machine is designed so that the colours in the terminal registers of $\lim_{s\to 1}M_s$ represent the solution to the computation.


\subsection{Single interaction adiabatic quantum machines}\label{SS:1xngaqc}

\begin{figure}
\centering
\psfrag{a}[c]{\small $H_0$}
\psfrag{b}[c]{\small $H_1$}
\psfrag{c}[l]{\small $H_{\mathrm{out}}$}
\includegraphics[width=0.15\textwidth]{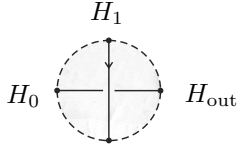}
\caption{\label{F:SingleCrossingAQC} An AQC with a single crossing.}
\end{figure}

The standard notion of adiabatic quantum computation corresponds to a machine with a single interaction, as pictured in Figure~\ref{F:SingleCrossingAQC}. A general AQC has multiple interactions, which we should consider as adiabatic computers working in conjunction to arrive at a solution. We will not details about adiabatic quantum machines in this paper--- we will only demonstrate what we have set out to: the way in which machine equivalence makes a difference in terms of computation. Our example involves only a single qubit.

Let $\sigma_x=\left(\begin{matrix}0 & 1\\
1 & 0\end{matrix}\right)$ and $\sigma_z=\left(\begin{matrix}1 & \phantom{-}0\\
0 & -1\end{matrix}\right)$ denote two out of three Pauli matrices, expressed with respect to the basis of $\mathds{C}^2$ consisting of the eigenvectors of $\sigma_z$. We use the standard notation, in which subscripts denote spin axes. Let $\mathds{1}$ denote the identity operator. Our adiabatic computer is designed to output the ground state $\bra 1\ket_z$. Choose the terminal Hamiltonian to be:
\begin{equation}
H_1\ass \frac{\mathds{1} + \sigma_z}{2} = \left(\begin{matrix}1 & 0 \\ 0 & 0\end{matrix}\right) = \bra 1\ket \kett 1\brat\enspace .
\end{equation}

\noindent Choose the initial Hamiltonian to be:

\begin{equation}
H_0\ass \frac{\mathds{1} - \sigma_z}{2} = \left(\begin{matrix}0 & 0 \\ 0 & 1\end{matrix}\right) = \bra 0\ket \kett 0\brat\enspace .
\end{equation}

\noindent The ground state of $H_0$ is $\bra 0 \ket_z$ and the ground state of $H_1$ is $\bra 1 \ket_z$.


At time $s$, our interaction has input $H_0$, agent $H_1$, and output
\begin{equation}
H_{\mathrm{out}}(s) = H_0 \trr H_1 = \frac{\mathds{1} + (2s-1) \sigma_z}{2}= \left(\begin{matrix}s & 0\\
0 & 1-s\end{matrix}\right)\enspace .
\end{equation}
Starting from $H_0$, the system evolves $H_{\mathrm{out}}(s)$ towards $H_1$ as $s$ approaches $1$. The computation turns out to be infeasible because the minimal energy gap along the evolution path vanishes, $g\left (H_{\mathrm{out}}(1/2) \right)=0$. This is due to the problem Hamiltonians $H_0$ and $H_1$ sharing the same eigenbasis, causing the energy levels to cross one another. We say that such a machine is (computationally) \emph{infeasible}.

\subsection{Multiple interaction adiabatic quantum machines}\label{SS:AQCmultiple}

The level crossing problem described in Section~\ref{S:AQC}\ref{SS:1xngaqc} can be avoided by extending the machine to include more than one interaction. Equivalent variants of the proposed AQC machine are given in Figure~\ref{fig:adiabaticm}. All machines have registers coloured $H_0$, $H_{\mathrm{out}}$, and $H_{\frac{1}{2}}$. These colours do not depend on $s$. Set $H_{\frac{1}{2}}\ass \sigma_x$. The terminal colour $H_{\mathrm{out}}$ has the following form:

\begin{figure}[htb]
\centering
\psfrag{a}[c]{\small $H_{\frac{1}{2}}$}
\psfrag{b}[c]{\small $H_0$}
\psfrag{c}[c]{\small $H_{1}$}
\psfrag{d}[c]{\small $H_{\frac{1}{2}}\trr H_1$}
\psfrag{e}[l]{\small $H^{\prime\prime}$}
\psfrag{e1}[c]{\small $H^\prime$}
\psfrag{f}[c]{\small $H_{\mathrm{out}}$}
\psfrag{g}[c]{\small $G$}
\psfrag{r}[c]{\small \emph{feasible}}
\psfrag{s}[c]{\small \emph{feasible}}
\psfrag{t}[c]{\small \emph{infeasible}}
\includegraphics[width=0.8\textwidth]{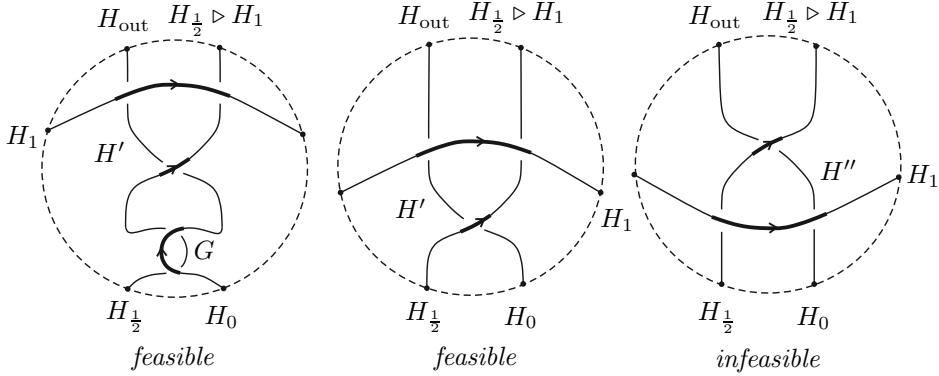}
\caption{\label{fig:adiabaticm}Equivalent adiabatic quantum machines.}
\end{figure}

\begin{equation}\label{E:R3AQC}
H_{\mathrm{out}} = (H_0 \trr \sigma_x) \trr H_{1} = (1-s)^2 H_0 + s(1-s) \sigma_x + s H_{1} = \begin{bmatrix}
s & s(1-s) \\
s(1-s) & (1-s)^2\rule{0pt}{13pt}
\end{bmatrix}\enspace.
\end{equation}

Thus if we write $H_{\mathrm{out}}(s)$ for $H_{\mathrm{out}}$ at time $s$, then $H_{\mathrm{out}}(s)\to H_0$ for $s\to 0$ while $H_{\mathrm{out}}(s)\to  \bra 1\ket \kett 1\brat= H_1$ for $s\to 1$. The ground state of $H_1$ is the sought-after solution.

The simple calculation of the classical adiabatic computer $H_0 \trr H_1$ in Section~\ref{S:AQC}\ref{SS:1xngaqc} has been replaced by the more involved computation of Equation~\ref{E:R3AQC}. We find that \begin{equation}g\left(H_{\mathrm{out}}(s) \right) = \left[ (s+(1-s))^2 - 4s(1-s)^3 \right]^{\frac{1}{2}}\enspace, \end{equation} \noindent and
$\min_s g\left( H_{\mathrm{out}}(s) \right) > \frac{2}{5}$. Thus we have solved the level crossing problem, and the final computations of each of the machines in Figure~\ref{fig:adiabaticm} are feasible.

A general AQC machine is fundamentally different from the single-crossing `classical adiabatic computation' in that it has intermediate stages at which
\emph{intermediate Hamiltonians} are present, describing neither the initial nor the terminal problem. 


As we are no longer interested only in the output $H_{\mathrm{out}}$ but also in the system as a whole, the Adiabatic Theorem should be applied also to all of the intermediate Hamiltonians in the AQC machine. With this in mind, let us examine the behavior of the intermediate Hamiltonians in the equivalent machines from Figure~\ref{fig:adiabaticm}.

The middle machine has two intermediate Hamiltonians that depend on $s$, namely, $\sigma_x \trr H_1$ and $H^\prime = H_0 \trr \sigma_x$, written
explicitly as:
\begin{equation}
\sigma_x\trr H_1(s) = \begin{bmatrix}
s & (1-s) \\
(1-s) & 0
\end{bmatrix} \quad\mbox{and}\quad
H^\prime(s) = \begin{bmatrix}
0 & s \\
s & (1-s)
\end{bmatrix}\enspace.
\end{equation}
\noindent Thus, $g\left(\sigma_x\trr H_1(s)\right) = \left[ s^2 + 4(1-s)^2 \right]^{\frac{1}{2}}$ and $g\left(H^\prime(s)\right) = \left[ (1-s)^2 + 4s^2 \right]^{\frac{1}{2}}$, both which have minimum energy gap $\min_s g \ge \frac{2}{\sqrt{5}}$. As the energy gaps $g(H^\prime)$ , $g(\sigma_x\trr H_1)$, and $g(H_{\mathrm{out}})$ in the middle machine are all non-vanishing throughout the adiabatic evolution, we conclude that this machine in its entirety represents a feasible computation.

Conversely, the machine on the right possesses no advantage compared to the classical adiabatic scheme. One of its Hamiltonians, $H^{\prime\prime} = H_0 \trr H_1$, has a vanishing energy gap for $s=\frac{1}{2}$. So taken as a whole, the machine on the right represents an infeasible computation.

The machine on the left in Figure~\ref{fig:adiabaticm} presents another equivalent feasible computation. Here the Hamiltonian $G = \sigma_x \rrt H_0$, where $\rrt$ is the inverse of the quandle operation $\trr$, has at least one negative eigenvalue for any $s \in [0,1)$,
\begin{equation}
G(s) = (1-s)^{-1}\left( \sigma_x - s H_0 \right) = \begin{bmatrix}
0 & (1-s)^{-1} \\
(1-s)^{-1} & -s(1-s)^{-1}
\end{bmatrix}\enspace .
\end{equation}

\section{Iteration and nesting}\label{S:Recursion}

Iteration lies at the heart of computational paradigms such as automata and Turing machines. It manifests the principle that the future state is determined exclusively by the current state and by subsequent inputs, via a \emph{transition function}. Similar concepts underlie several widely used probabilistic models such as Markov chains and autoregressive processes.

To realize iteration in a machine, consider copies $M_0,M_1,M_2,\ldots$ of a fixed machine $M$. As we are going to construct a new machine by concatenating these, we assume each of these copies to have its endpoints lying on a circular firmament. Partition these sets of `endpoint registers' into two subsets of the same size $\textrm{In}(M_i)$ and $\textrm{Out}(M_i)$, and associate a unique \emph{terminal register} in $\textrm{Out}(M_i)$ to each \emph{initial register} in $\textrm{In}(M_i)$. We graphically indicate an element of $\textrm{In}(M_i)$ with an arrow from the firmament into the disk it bounds, and an element of $\textrm{Out}(M_i)$ with an arrow out of the disk to the firmament. We refer to registers of closed processes in $M$ as \emph{control registers}. Write $\mathrm{U}(M)$ for the set of control registers in $M$.

Initialize the registers of $\textrm{In}(M_0)$ to the initial state of the iteration, and initialize also the control registers within each $M_i$. Concatenate each terminal register in $M_i$ with its corresponding initial register in $M_{i+1}$ for $i=0,1,2,\ldots$. Denote the resulting machine $\tilde{M}$. For each $i=0,1,2,\ldots$, the result of the computation of $M_i$ appears as the colours stored in $\textrm{Out}(M_i)$, assuming these are uniquely determined by $\textrm{In}(M_i)$ and by $\mathrm{U}(M_i)$. Given the initial condition and colours for the control registers, the computation of $\tilde{M}$ is its \emph{steady state}, that is the set of colours in $\textrm{In}(M_N)$ where $N\geq 0$ is such that each initial register has the same colour as its corresponding terminal register in $M_n$ for all $n>N$. A steady state can be diagrammatically described via a colouring of the \emph{closure} of $M$ (concatenating each terminal vertex with its corresponding initial vertex). Conversely, $M$ may compute the set of initial conditions for which a steady state exists.

Special cases of the above computational paradigm have been studied in \citep{Kauffman:94, Kauffman:95}. His tangles consist of a single open process, and the iteration represents feedback loops which are a research interest of Kauffman and a primary ingredient in cybernetic sciences. Using a quandle colouring, Kauffman showed that such \emph{long knots} underlie a class of automata which can emulate multi-valued logic and modular arithmetic computations. An example he considers is based on iterating a `trefoil machine' $M$ in which initial registers $x_0$ and $y_0$ are coloured $a$ and $b$ in some quandle $Q$ whose underlying set underlies a field $F$ and whose operation is $a\trr b= 2b-a$. The iteration machine $\tilde{M}$ attains a steady state if and only if $3(a-b)=0$, \textit{i.e.} if and only if $a-b$ is an element of order $3$ in $F$.

\begin{figure}
\centering
\psfrag{a}[c]{\small $u_0$}
\psfrag{b}[c]{\small $u_1$}
\psfrag{c}[c]{\small $u_2$}
\psfrag{d}[c]{\small $u_3$}
\psfrag{e}[c]{\small $u_4$}
\psfrag{f}[c]{\small $u_n$}
\psfrag{g}[c]{\small $x_{0:n}$}
\psfrag{h}[c]{\small $x_{1:n}$}
\psfrag{j}[c]{\small $x_{2:n}$}
\includegraphics[width=0.75\textwidth]{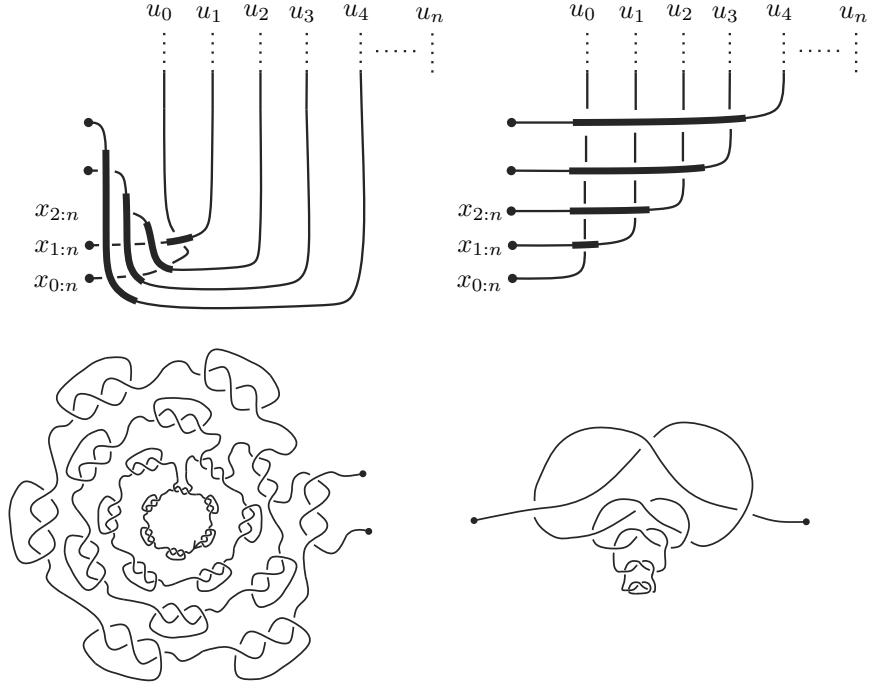}
\caption{\label{F:Nesting}Nested machines}
\end{figure}

Figure~\ref{F:Nesting} shows some examples of iteration, and of a more general construction which we call \emph{nesting}. The machines on the upper row are studied in Section~\ref{S:Recursion}\ref{SS:SLRecursion}, while the remainder of the section considers a machine which models a Markov chain.

\begin{rem}
Our theory does not account for machines with infinitely many interactions, so we may assume that the nesting is large but finite. This assumption has nothing to do with whether or not the process halts, whatever halting means in our context.
\end{rem}

\subsection{Basic linear iteration}\label{SS:SLRecursion}

We investigate the computation of the equivalent machines in the upper row of Figure~\ref{F:Nesting}. We colour these machines by a linear quandle (Example~\ref{E:LinearQuandle}) which has only a single operation $x\trr y \ass (1-s)x+sy$ for some fixed $s\in \mathds{R}\setminus \{1\}$ (and its inverse $\rrt$). 
The initial colours in both machines are given as $u_0,u_1,\ldots \in \mathds{R}$. The terminal colours are then computed to be:
\begin{equation}\label{E:xon}
x_{0:n} = (1-s)^n u_0 + s \sum_{i=0}^{n-1} (1-s)^i u_{n-i}\enspace,
\end{equation}
\noindent which can be expressed concisely as
%
$x_{0:i} = x_{0:i-1} \trr u_i$
%
with $x_{0:0} = u_0$. This iteration describes a dynamical system, or more precisely, an equivalence class of such systems whose behavior is dictated by the fixed \emph{quandle parameter} $s$ and by the inputs $u_0,u_1,\ldots$.

Equation~\ref{E:xon} expresses $x_{0:n}$ as a sum of an effect of the initial condition $u_0$ with a discrete-time convolution of $(1-s)^i$ with the inputs $u_i$, where $i$ is the discrete-time index. This expression may be viewed as a generating function encoding information about the inputs, by rewriting it as:
\begin{equation}
x_{0:n} = \sum_{i=0}^n w_i(s) u_{n-i}\enspace .
\end{equation}
The coefficients $w_i(s)$, $i=0,1,2,\ldots$ all are machine invariants, \textit{i.e.} for equivalent machines they are the same.

Aside from $x_{0:n}$, the machines also compute $x_{1:n}, \ldots, x_{n:n}$. These all are the outputs of related dynamical systems with increasingly smaller evolution histories. Thus $x_{k:n}$ is the output of a system whose initial state is $x_{k:k}=u_k$ and which so far has processed $n-k$ inputs.

\subsection{Markovian links}\label{SS:MarkovMachine}

We next present a more involved example of an iterative computation.


\begin{figure}[htb]
\centering
\psfrag{a}[c]{\small $v_i^1$}
\psfrag{b}[c]{\small $v_i^2$}
\psfrag{c}[l]{\small $v_{i+1}^1$}
\psfrag{d}[c]{\small $v_{i+1}^2$}
\psfrag{e}[c]{\small $\trr_1$}
\psfrag{f}[c]{\small $\trr_2$}
\psfrag{m}[c]{\small \emph{sub-machine} $M_i$}
\psfrag{m1}[c]{\small $M_{i-1}$}
\psfrag{m2}[c]{\small $M_i$}
\psfrag{m3}[c]{\small $M_{i+1}$}
\psfrag{u}[l]{\small $\pi^1$}
\psfrag{v}[c]{\small $\pi^2$}
\psfrag{x}[c]{\small $v_{i-1}^2$}
\psfrag{y}[c]{\small $v_{i-1}^1$}
\psfrag{x1}[l]{\small $v_{i+2}^1$}
\psfrag{y1}[l]{\small $v_{i+2}^2$}
\psfrag{s}[c]{\small \emph{coloured closure (stationarity)}}
\psfrag{t}[c]{\small \emph{iteration} $\tilde{M}$}
\includegraphics[width=0.93\textwidth]{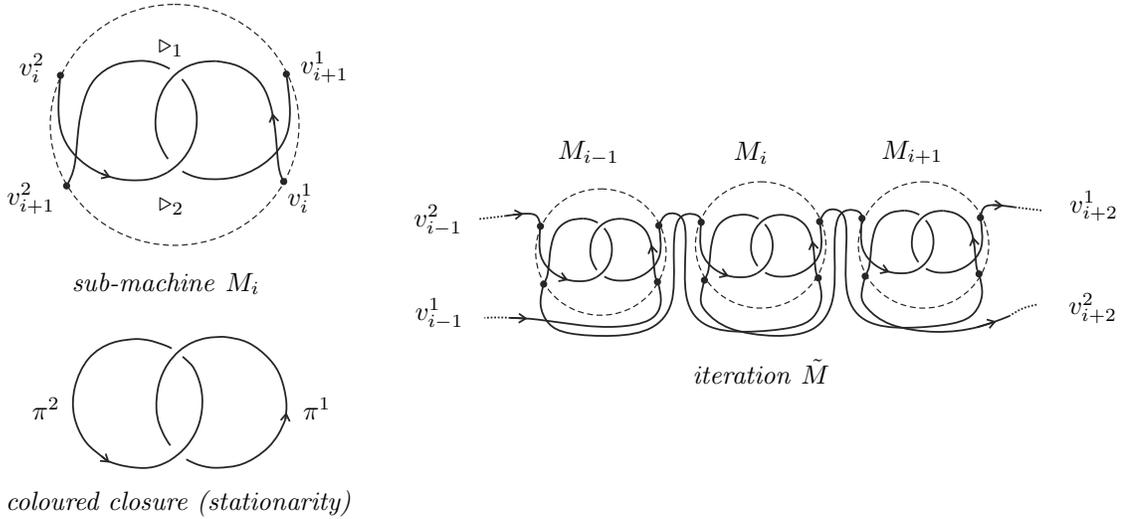}
\caption{\label{fig:hopfm} An iterative machine and its steady-state.}
\end{figure}

Consider the iteration machine $\tilde{M}$ built out of concatenating identical copies $M_0$, $M_1$, $M_2$, \ldots of the machine pictured in the upper left corner in Figure~\ref{fig:hopfm} by concatenating the two terminal registers $v^1_{i+1}$ and $v^2_{i+1}$ in $M_i$ to their namesake initial registers in $M_{i+1}$. For this example, consider a linear quandle with two operations $\trr_{s_1}$ and $\trr_{s_2}$ with $s_1,s_2\neq 1$. To avoid degenerate cases we assume also that $s_1,s_2\neq 0$. We abbreviate the names of these operations to $\trr_1$ and $\trr_2$ correspondingly. 

For the specified concatenation to be defined, the following relation between $\mathrm{Out}(M_i)$ and $\mathrm{In}(M_i)$ must be satisfied:
\begin{equation}
\label{eq:hopfm}
\left\{v_{i+1}^1 = v_i^1 \trr_2 v_i^2, \; v_{i+1}^2 = v_i^2 \trr_1 v_i^1 \right\} \; \; \longrightarrow \; \;
v_{i+1}  = \underbrace{\begin{bmatrix}
1-s_2 & s_2 \\
s_1 & 1-s_1
\end{bmatrix}}_P v_i\enspace ,
\end{equation}
\noindent where $v_i \ass \left[\begin{matrix}v_i^1\\ \rule{0pt}{13pt}v_i^2\end{matrix}\right]$.

All entries of the \emph{one-step transition matrix} $P$ are non-negative, and each of its rows sums to $1$. A matrix with such properties is said to be \emph{(right) stochastic}.

The Perron--Frobenius Theorem for stochastic matrices tells us that $P$ has a unique largest eigenvalue equal to $1$ whose corresponding eigenvector $\pi$ has strictly positive entries. Again by Perron--Frobenius, for \emph{any} vector $v_0$ of probabilities satisfying $\sum_j v_0^j =1$, the homogenous irreducible Markov chain with one-step transition matrix $P$ converges to $\pi$ irrespective of the initial distribution $v_0$:

\begin{equation}\lim_{i \to \infty} P^i v_0 = \pi\enspace.\end{equation}


The iteration machine $\tilde{M}$ represents an homogeneous irreducible Markov chain whose one-step transition matrix $P$ is given by \eqref{eq:hopfm}. We have shown that $\tilde{M}$ has a steady-state, which we may describe by `closing' a machine $M_i$:
\begin{equation}
\mathrm{In}(M_i) = \mathrm{Out}(M_i) \; \; \longrightarrow \; \; \pi = P \pi\enspace .
\end{equation}
Thus, $\pi=\left[\begin{matrix}\pi^1 \\ \pi^2\end{matrix}\right]$ is the eigenvector of $P$ corresponding to the eigenvalue~$1$.

\begin{rem}
In the special case $s_1=s_2$, matrix $P$ is \emph{doubly stochastic}.
\end{rem}

\subsection{Feed-forward}\label{SS:FeedForward}

Figure~\ref{fig:hopfm} depicts a machine analogous to an homogeneous irreducible Markov chain, for which a steady-state colouring is always attained. Such machines are said to be \emph{(externally) stable}. The machine $\tilde{M}$ is also \emph{internally stable}, meaning that for any concatenation of machines that gives rise to $\tilde{M}$, each transition matrix describing the concatenation is stochastic.

In this section and the next, we shall exhibit equivalent machines to $\tilde{M}$ which are not internally stable. To the best of our knowledge, there is no competing formalism in the literature for which to discuss equivalent Markov chains which may or may not be internally stable.

\begin{figure}[htb]
\centering
\psfrag{m1}[c]{\small $M_i$}
\psfrag{m2}[c]{\small $M_{i+1}$}
\psfrag{x}[c]{\small $v_i^2$}
\psfrag{y}[c]{\small $v_i^1$}
\psfrag{u}[l]{\small $v_{i+1}^2$}
\psfrag{v}[l]{\small $v_{i+1}^1$}
\psfrag{s}[l]{\small $v_{i+2}^2$}
\psfrag{r}[l]{\small $v_{i+2}^1$}
\psfrag{t}[c]{\small \emph{feed-forward machine} $M^\prime$}
\psfrag{a}[c]{\small \emph{feed-back machine} $M^{\prime\prime}$}
\includegraphics[width=0.6\textwidth]{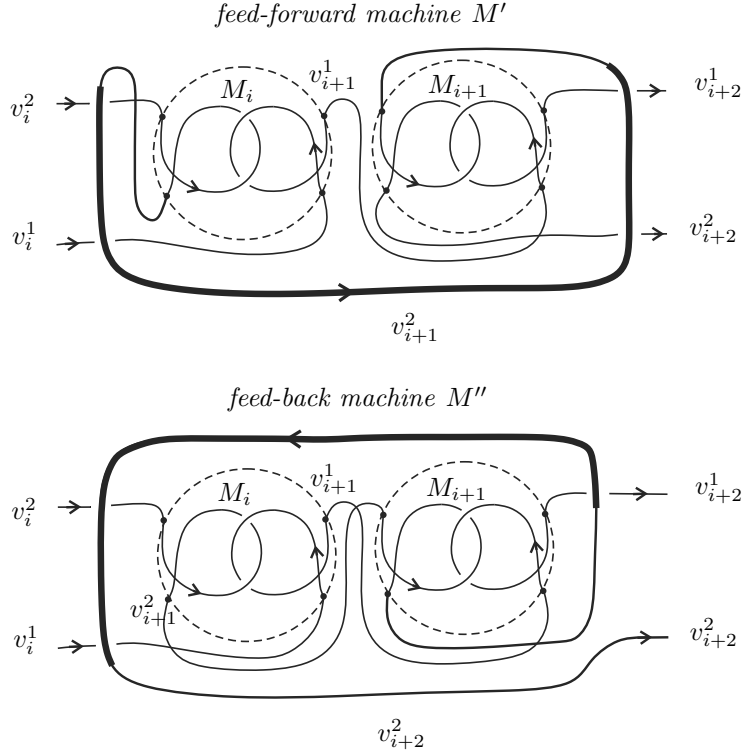}
\caption{\label{fig:hopfm1} Feed-forward and feed-back equivalent machines.}
\end{figure}

Consider a machine $\tilde{M}^\prime \sim \tilde{M}$ built from concatenating (`stacking') copies $M_0^\prime, M_1^\prime,\ldots$ of the upper machine $M^\prime$ in Figure~\ref{fig:hopfm1} by concatenating each register in $M_i^\prime$ with its namesake register in $M_{i+1}^\prime$. The machines are now coloured by a linear quandle with three operations (and their inverses) corresponding to three real numbers $s_1,s_2,s_3\neq 0,1$. The `feed-forward machine' $M^\prime$ is created by sliding the concatenated output strand $v_{i+1}^2$ of $M_i$ all the way across the outputs of $M_{i+1}$, crossing over the inputs of $M_i$. This overcrossing strand, pictured as a thickened line, acts as an agent via $\trr_3\ass \trr_{s_3}$. Metaphorically, we are using a colour $v_{i+1}^2$ `from the past' to manipulate colours $v_{i+2}^1$ and $v_{i+2}^2$ `in the future'. 

By the equivalence of $M$ with $M^\prime$, we know that:
\begin{equation}
\label{eq:sym}
v_{i+2} = P^2 v_i
\end{equation}
where, as before, $v_i=\left[\begin{matrix}v_i^1\\ \rule{0pt}{13pt}v_i^2\end{matrix}\right]$. Unlike in $\tilde{M}$, the colours $v_{i+1}$ and of $P v_i$ need not coincide in $\tilde{M}^\prime$. 
Writing $P_i$ for the matrix such that $v_{i+1}=P_i v_i$, instead of the relation $v_{i+1}=Pv_i$ for $\tilde{M}$, we now obtain the pair of relations $v_{2i} = P_{1} v_{2i-1}$ and $v_{2i+1} = P_0 v_{2i}$, where $P_{0} P_1 = P^2$. Thus, the one step transition matrices in $\tilde{M}$ all equal $P$, while in $\tilde{M}^\prime$ the transition matrix from $v_{n}$ to $v_{n+1}$ is $P_{n\bmod 2}$.

We compute $P_0$ and $P_1$ explicitly:
\begin{subequations}
\begin{equation}
P_0 = \begin{bmatrix}
(1-s_2-s_1 s_3)(1-s_3)^{-1} & (s_2-s_3+s_1 s_3)(1-s_3)^{-1} \\
s_1 & 1-s_1
\end{bmatrix}\enspace ;
\end{equation}
\begin{equation}
P_{1} = \begin{bmatrix}
(1-s_2)(1-s_3) & s_2(1-s_3)+s_3 \\
s_1(1-s_3)\rule{0pt}{13pt} & (1-s_1)(1-s_3)+s_3
\end{bmatrix}\enspace .
\end{equation}
\end{subequations}

The important point is that $P_0$ and $P_1$ may no longer be stochastic as some of their entries may be negative or greater than $1$. But $P_0P_1=P^2$ is stochastic and well-behaved. So perhaps the computation of internal colours in each $M_j^\prime$ should be thought of as \emph{abstract}. Moreover, the internal colours of $v_{i+1}$ registers may not even be bounded for $s_3$ sufficiently close to $1$, and therefore they may not represent probabilities. Thus, $M^\prime$ and $\tilde{M}^\prime$ are internally unstable.

\subsection{Feed-back}\label{SS:Feedback}

Next consider the \emph{feed-back machine} $M^{\prime\prime}$ in Figure~\ref{fig:hopfm1}. It is formed by sliding the output strand $v_{i+2}^2$ all the way back across the inputs of $M_i$. It is as though a `future' register manipulates `past' ones. Similarly to the feed-forward machine, the feed-back machine may be internally unstable. Its structure is yet more intricate in that it resembles a regulating control loop such as those which are encountered in the theory of dynamical systems and in cybernetics.

We again compute the relations between $\mathrm{Out}(M^{\prime\prime})$ and $\mathrm{In}(M^{\prime\prime})$ required for concatenation.
For the feed-back machine, $v_{2i} = P_{1} v_{2i-1}$ with:
\begin{equation}
P_{1} = \begin{bmatrix}
(1-s_2)(1-s_3) & s_2(1-s_3)+s_3 \\
s_1(1-s_3) & (1-s_1)(1-s_3)+s_3\rule{0pt}{13pt}
\end{bmatrix}\enspace .
\end{equation}
For the transition from $v_{2i}$ to $v_{2i+1}$, we compute:
\begin{equation}
v_{2i+1} =
\underbrace{
\begin{bmatrix}
(1-s_2)(1-s_3)^{-1} & s_2(1-s_3)^{-1} \\
s_1(1-s_3)^{-1} & (1-s_1)(1-s_3)^{-1}\rule{0pt}{13pt}
\end{bmatrix}}_{P_0^{\prime\prime}} v_i +
\underbrace{
\begin{bmatrix}
0 & -s_3 (1-s_3)^{-1}  \\
0 & -s_3 (1-s_3)^{-1}\rule{0pt}{13pt}
\end{bmatrix}}_{T} v_{2i}\enspace .
\end{equation}

We deduce that:
\begin{equation}
\label{eq:inout}
v_{2i+1} = \left(P_0^{\prime\prime} + T P^2\right) v_{2i}, \quad v_{2i+2} = P_{1} \left(P_0^{\prime\prime} + T P^2\right) v_{2i}\enspace .
\end{equation}

\noindent Moreover, \eqref{eq:sym} attests that $P^2$ equals $P_{1} \left(P_0^{\prime\prime} + T P^2\right)$, because both map $v_i$ to $v_{i+2}$. Hence we find that:
\[
P^2 = \left(I - P_{1} T \right)^{-1} P_{1} P_0^{\prime\prime}\enspace ,
\]
\noindent which leads to
\begin{equation}
\label{eq:inout1}
v_{2i+1} = \left( I - P_{1} T \right)^{-1} P_{1} P_0^{\prime\prime} v_{2i}\enspace .
\end{equation}

Relations of the form \eqref{eq:inout1} are encountered in the theory of dynamical systems, where they manifest a regulating procedure known as a closed (control) loop. In the context of machines, a closed loop is interpreted as follows. Any machine equivalent to the internally stable machine $\tilde{M}$ is stable, but not necessarily internally stable. We might imagine islands of instability in an externally stable cosmos (machine). Feed-back and feed-forward machines which are not internally stable regulate their behavior so as to become externally stable. In our example, the one-step transition matrices are not stochastic, but the two-step transition matrices are stochastic.

\section{Conclusion}

We have introduced tangle machines as a diagrammatic algebra uniting ideas in low-dimensional topology, causality, information, and computation. There is a natural local notion of tangle machine equivalence. We have exhibited ways in which machine equivalence may represent networks with identical global properties, but with different local properties, within a number of different paradigms of computation. Our vision is to model these and other complex real-world phenomena by machines, then to use machine equivalence to select a `best' machine (whatever `best' means in that context), and then to perform a computation for that `best' machine which might not have been tractable for the machine that we started with.

Future work will discuss topological invariants of machines, will expand on our examples, will discuss statistical detection of machines inside data, and will discuss algorithmic aspects of finding a `best' machine inside an equivalence class.

\bibliographystyle{rspublicnat}

\newpage

\appendix
\section*{Appendix}

\subsection{Formal definition of tangle machines}\label{A:TangleMachines}

We present a rigourous definition of tangle machines, and show how it is equivalent to the result of the constructions of Section~\ref{S:Machines}.

\begin{defn*}[Tangle machines]\label{D:TangleMachine}
A \emph{tangle machine} $M$ coloured by a quandle $(Q,B)$ is a quintuple $M\ass (G,S,\phi,\varrho,\rho)$ consisting of:
\begin{itemize}
\item A finite graph $G$ that is a disjoint union of path graphs $P_1,\ldots, P_k$  and cycles $C_1,\ldots,C_l$:
\begin{equation}
G\,\ass\, \left(P_1\dU P_2\dU \cdots\dU P_k\right)\dU \left(C_1\dU C_2\dU\cdots\dU C_l\right),
\end{equation}
The graph $G$ is called the \emph{underlying graph} of $M$. Vertices of $G$ are called \emph{registers}.
\item A subset of registers $S\subseteq V(G)$ called \emph{agents}.
\item A multivalued \emph{interaction function} $\phi\colon\, S\Rightarrow E(G)$ specifying the edges acted on by each agent.
\item An \emph{operation function} $\varrho\colon\, S\to B$ specifying the action of each agent.
\item A \emph{colouring function} $\rho\colon\,V(G)\to Q$ such that if $v$ and $w$ are registers in $M$ and if $e$ is an edge from $v$ to $w$ then $\rho(v)=\rho(w)$ if $e\notin \mathrm{Im}(\phi)$. Otherwise let $u\in S$ be the vertex such that $\phi(u)=e$ and set $\trr \ass \varrho(u)$. Then either $\rho(v)\trr \rho(u)=\rho(w)$ or $\rho(w)\trr \rho(u)=\rho(v)$.
\end{itemize}
\end{defn*}

 To draw a tangle machine, first draw the graph $G$, then draw a dotted line between each agent $u$ in $S$ and the edges in its image, with the $\varrho(u)$ indicated on each of these edges. Finally, label each register by its image under $\rho$.

Reidemeister moves are defined as follows:

\begin{description}
\item[Reidemeister \textrm{I}:] For $(x,\trr)\in (Q,B)$:
\begin{equation}\label{E:R1}
%
\includegraphics[width=0.85\linewidth]{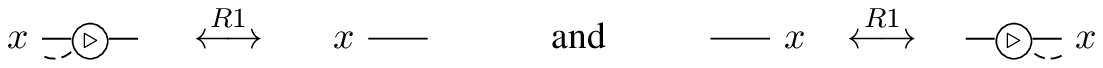}
\end{equation}



\item[Reidemeister \textrm{II}:] In following local modification, the top central register must be outside the set of agents $S$. Here, $x,y\in Q$ and $\trr\in B$.
\begin{equation}
\includegraphics[width=0.99\linewidth]{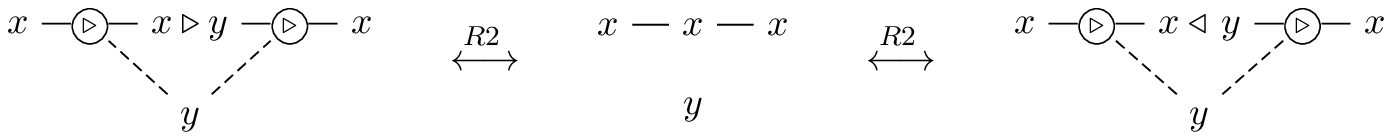}
\end{equation}
%
%
%
\item[Reidemeister \textrm{III}:] Here, $x,y,z\in Q$ and $\trr,\brr\in B$. Writing $u$ for the register coloured $y$ in the move below, all edges in $\phi(u)$ must participate in the move (the move is invalid for a strict subset of them):

\begin{equation}
\includegraphics[width=0.9\linewidth]{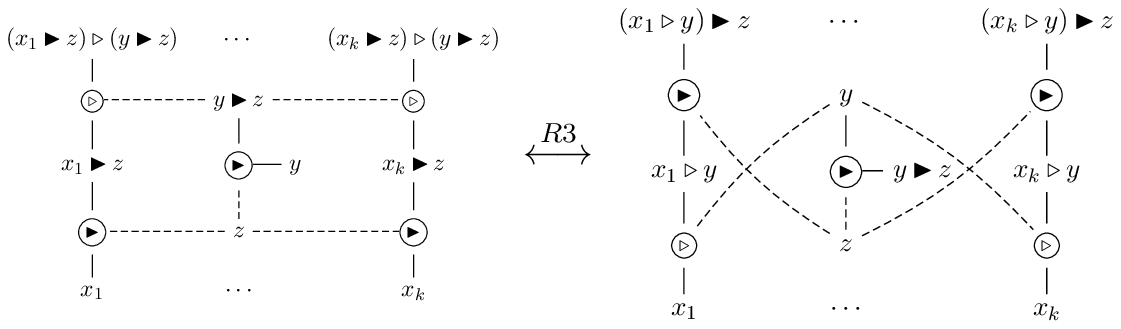}
\end{equation}


For example, one R3 move for $k=0$ reads:

\begin{equation}
\includegraphics[width=0.7\linewidth]{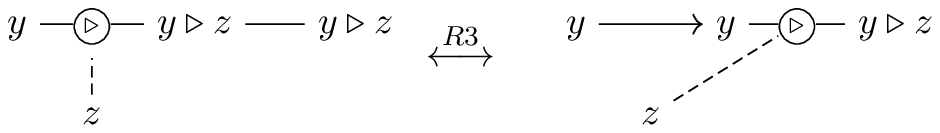}
\end{equation}

One R3 move for $k=1$ reads:

\begin{equation}
\includegraphics[width=0.95\linewidth]{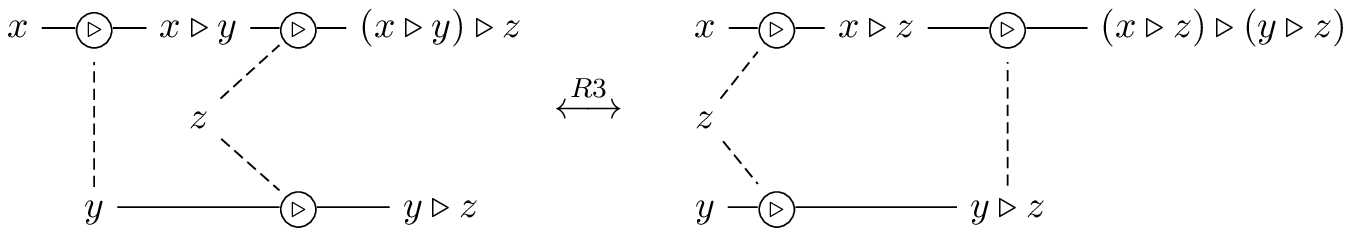}
\end{equation}
\end{description}

We define also a \emph{stabilization}, where $x\in Q$:

\begin{equation}
\includegraphics[width=0.2\linewidth]{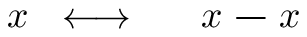}
\end{equation}

\begin{defn*}
Two tangle machines are \emph{equivalent} if they are related by an automorphism of $(Q,B)$ followed by a finite number of Reidemeister moves. The machines $M_{1,2}$ are \emph{stably equivalent} if there exist equivalent machines $M_{1,2}^\prime$ such that $M_1^\prime$ is obtained from $M_1$ by a finite sequence of stabilizations and $M_2^\prime$ is obtained from $M_2$ by a finite sequence of stabilizations.
\end{defn*}

Equivalence of tangle diagram descriptions is, on the other hand, defined as follows:

\begin{defn*}
Tangle diagrams $M$ and $M^\prime$ of tangle machines are considered \emph{equivalent} if they (or rather their restrictions to a closed disk outside which they both consist only of rays to infinity) are related by an automorphism of $(Q,B)$ together with planar isotopies and a finite sequence of cosmetic moves  (Figure~\ref{F:local_moves_machines}) and Reidemeister moves (Figure~\ref{F:local_moves_machines1}).
\end{defn*}

To obtain our `tangle description' from the above definition, first destabilize until each edge is in the $\phi$--image of some agent. Then replace each `interaction' (an agent in $S$ together with all edges in its $\phi$--image) by an `interaction' in the sense of Section~\ref{S:Machines}\ref{SS:TangleMachines}:

\begin{equation}\label{E:kebab}
\includegraphics[width=0.85\linewidth]{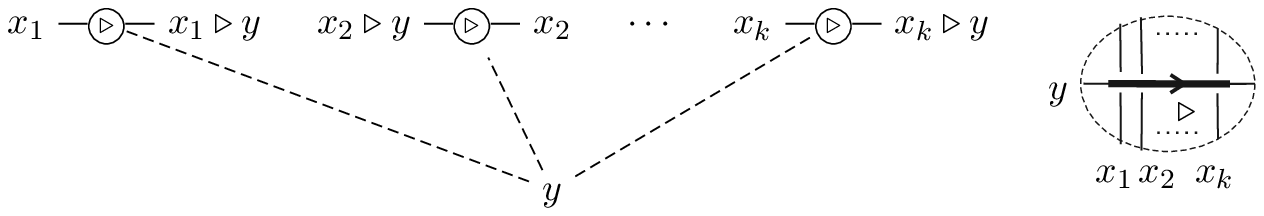}
\end{equation}

The indeterminacy in the translation from Gau{\ss} diagram interactions to tangle diagram interactions is captured by moves $I1$, $I2$ and $I3$ in Figure~\ref{F:local_moves_machines}. 

Concatenate as prescribed by the graph. The indeterminacy in doing this is captured by moves $V\!R1$, $V\!R2$, $V\!R3$, and $SV$ in Figure~\ref{F:local_moves_machines}. Once tangle endpoints have been `sent to infinity', there are no further indeterminacies. Reidemeister moves on quintuples $(G,S,\phi,\varrho,\rho)$ correspond to the Reidemeister moves of Figure~\ref{F:local_moves_machines1} by construction.

A quandruple $(G,S,\phi,\varrho,\rho)$ may be considered as a \emph{Gau{\ss} diagram} of its corresponding tangle diagram.
To translate from a tangle diagram back to a Gau{\ss} diagram, get rid of interactions without patients using $ST$ and reverse the above process. There are no indeterminacies. We have proven the following proposition:

\begin{prop*}
Stable equivalence classes of Gau{\ss} diagrams of machines are in bijective correspondence with equivalence classes of tangle diagrams of machines.
\end{prop*}

\begin{rem*}
In fact $ST$ (whose left-hand side may result from an $R2$ move) should be thought of as a stabilization and equivalent tangle diagrams should be called \emph{stably equivalent}. We adopt the present convention for main-text simplicity.
\end{rem*}

We conclude with two examples of Gau{\ss} diagrams and corresponding tangle diagrams. The first example features a quandle for which $a\trr (b\trr a)=(a\trr b)\trr a = b$ for all $a,b\in Q$ and for all $\trr\in B$. In the second example, colours are suppressed.

\begin{equation}
\includegraphics[width=0.85\linewidth]{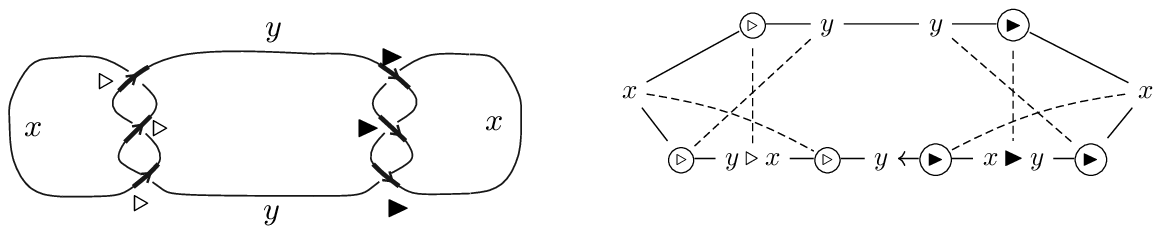}
\end{equation}

\begin{equation}
\includegraphics[width=0.85\linewidth]{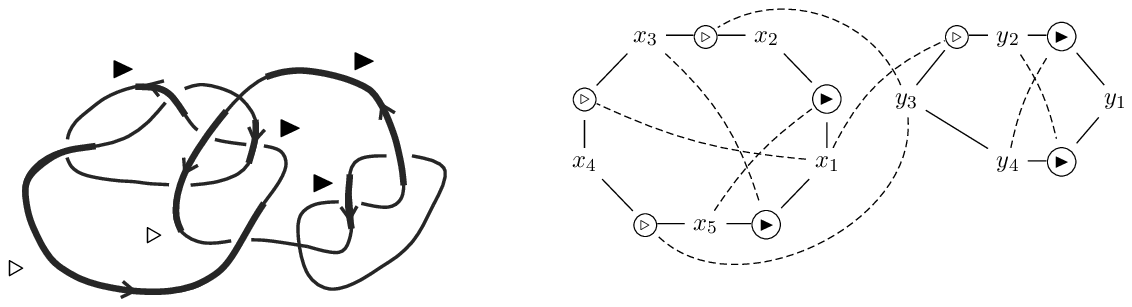}
\end{equation}

\end{document}